\documentclass[draftcls,onecolumn,11pt]{IEEEtran}
\usepackage{ifpdf}

\usepackage{cite}

\ifCLASSINFOpdf
\usepackage[pdftex]{graphicx}
\DeclareGraphicsExtensions{.pdf,.jpeg,.png} \else
\usepackage[dvips]{graphicx}
\DeclareGraphicsExtensions{.eps} \fi
\usepackage[cmex10]{amsmath}
\usepackage{algorithm}
\usepackage{algorithmic}

\usepackage{array}

\usepackage{mdwmath}
\usepackage{mdwtab}

\usepackage{eqparbox}

\usepackage[tight,footnotesize]{subfigure}

\usepackage{stfloats}
\usepackage{amssymb}

\renewcommand{\IEEEQED}{\IEEEQEDopen}

\begin{document}
%
\title{A Multi-Interference-Channel Matrix Pair Beamformer for CDMA Systems}
%
%
%

\author{Jian~Wang, Jianshu~Chen,~\IEEEmembership{Student Member,~IEEE,} ~Jian~Yuan, \\
Ning Ge,~\IEEEmembership{Member,~IEEE} and ~Shuangqing Wei, ~\IEEEmembership{Member,~IEEE}
\thanks{
                This work was supported by the National Natural Science Foundation of China under Grant No. 60928001 and No. 60972019.
            }
        \thanks{
                Jian Wang, Jian Yuan and Ning Ge are with the Department of Electronic Engineering,
                Tsinghua University, Beijing, P. R. China, 100084. (e-mail: \{jian-wang, jyuan, gening\}@tsinghua.edu.cn)
               }
        \thanks{
                Jianshu Chen is with the Department of Electrical Engineering, University of
                California, Los Angeles, CA 90095-1594, USA. (e-mail: cjs09@ucla.edu)
                }
        \thanks{
                Shuangqing Wei is with the Department of Electrical and Computer Engineering, Louisiana State University, Baton Rouge, LA 70803, USA. (e-mail: swei@ece.lsu.edu). His work was supported in part by Louisiana Board of Regents under Grant No. LEQSF(2009-11)-RD-B-03 and the National Science Foundation (NSF) of US under Grant No. CNS-1018273.
                }
}
\maketitle

\begin{abstract}
Matrix pair beamformer (MPB) is a promising blind beamformer
which exploits the temporal signature of the signal of interest
(SOI) to acquire its spatial statistical information. It does not
need any knowledge of directional information or training sequences.
However, the major problem of the existing MPBs is that
they have serious threshold effects and the thresholds will grow as
the interference power increases or even approach infinity. In
particular, this issue prevails in scenarios with structured
interference, such as, periodically repeated white noise, tones, or
MAIs in multipath channels. In this paper, we will first present the
principles for designing the projection space of the MPB
which are closely correlated with the ability of suppressing
structured interference and system finite sample performance. Then a
multiple-interference-channel based matrix pair beamformer (MIC-MPB)
for CDMA systems is developed according to the principles. In order
to adapt to dynamic channels, an adaptive algorithm for the
beamformer is also proposed. Theoretical analysis and simulation
results show that the proposed beamformer has a small and bounded
threshold when the interference power increases. Performance
comparisons of the MIC-MPB and the existing MPBs in
various scenarios via a number of numerical examples are also
presented.
\end{abstract}


%
\IEEEpeerreviewmaketitle

\section{Introduction}
\label{Sec:Introduction} 

Adaptive beamforming is a promising technique to spatially suppress
interference, and can be used in dense interference environments,
such as, direct sequence code division multiple access (DS-CDMA)
systems. Adaptive beamforming techniques often make use of a known
training sequence or the direction-of-arrival (DOA). However, the
time-varying nature of mobile communication requires continuous DOA
tracking or pilot signals in these methods, which increases the
complexity and bandwidth requirement. In addition, steering vector
errors will cause performance loss in DOA-based beamformers as
well\cite{Trees2002,WaxApr1996}.

To overcome these problems, many blind adaptive beamforming
algorithms have been extensively studied. The constant modulus
algorithm (CMA) is a class of gradient-based algorithm that works on
the premise that the existence of an interference causes fluctuation
in the amplitude of the array output, which otherwise has a constant
modulus \cite{Godara1997,Jani2000,Paik2006,Lamare2008}. But for the
possible presence of constant modulus (CM) interfering signals (e.g.
MAI, BPSK jamming, etc.) and the requirement for power control, the
blind algorithm based on CM property is less feasible for DS-CDMA
systems \cite{torrieri2007maa}. Another class of blind algorithms
exploit the temporal signature of the signal of interest (SOI) to
acquire its spatial statistical information, which also only
requires the spreading code and timings of the desired user
\cite{Naguib1996,Song2001,choi2002nab,Yang2006fbba,torrieri2004,torrieri2007maa}
as the CMA methods \cite{Jani2000,Paik2006,Lamare2008}. In
\cite{Naguib1996,Song2001,choi2002nab,Yang2006fbba}, the
eigenstructures of the pre- and post-correlation (PAPC) array
covariance matrices are used to derive the beamformer, and various
kinds of low complexity iteration algorithms are developed. The
Maximin algorithm proposed in \cite{torrieri2004,torrieri2007maa}
uses a filter pair (FP) to separate the SOI and the interference,
and update the weight vector by steepest decent method.

As indicated in our recent work \cite{Jianshu,Jianshu2}, these
approaches share the same processing structure, i.e., two
projections to construct two estimated matrices followed by a
generalized eigen-decomposition of the matrix pair, and hence are
referred to as matrix pair beamformer (MPB). We also find the key
assumption that the two matrices share the same interference
statistics is not valid in many cases, which will cause so-called matrix mismatch \cite{Jianshu,Jianshu2}. Due to matrix
mismatch, the MPB always suffers from a threshold effect.
When the input signal-to-noise ratio (SNR) is below the threshold,
the performance of the beamformer will degrade rapidly, and the main
beam will point to the direction of interferers. In some cases, the
threshold SNR is infinity and the MPB fails forever.
Furthermore, the existing MPB is vulnerable to structured
interference in many cases, such as periodically repeated white
noise, tones, and MAIs in multipath channels. As a result, the
threshold will grow as the interference power increases. In order to
make the beamformer work, the power of the SOI should also increase
to compete with that of the interference. This property means the MPB cannot function under this condition. Therefore, it is
important to design an MPB with ability of suppressing
structured interference.

Finite sample effect is another important factor having an impact on
the performance of a beamformer. Since insufficient sample-support
may cause a considerable mismatch between true and sample covariance
matrices in practical implementations, the calculated noise
eigenvalues will be a significant spread around the correct values
\cite{Trees2002}. As a result, how much independent noise samples
obtained can determine the performance of a beamformer. Robust
design of a beamformer involving diagonal loading factor
\cite{Jianli2005,Mestre2006} is another approach to cope with this
problem, which desensitizes the system by compressing the noise
eigenvalues of the correlation matrix so that the nulling capability
against small interference sources is reduced \cite{Mestre2006}.
However, how to choose the best loading factor in a real scenario in
order to combat the finite sample effect is still an open problem.

Based on the above observations and the analytic results in our
recent work, in this paper, we first propose several principles for
designing the projection space for MPBs. Then a
multiple-interference-channel based matrix pair beamformer (MIC-MPB)
for CDMA systems is developed. The beamformer has a small and
bounded threshold, i.e., the threshold does not grow when the power
of the interference increases. Moveover, by exploiting more
signal-free interference samples, the approach achieves a less
perturbed noise subspace and avoids signal cancelation.

The rest of the paper is organized as follows. Section II presents a
general framework of MPB to summarize and reinterpret the
basic ideas in
\cite{Naguib1996,Song2001,choi2002nab,Yang2006fbba,torrieri2004,torrieri2007maa},
followed by reviewing some results concerning threshold effects of
the existing MPBs. In Section III, we first present the
principles for designing the projection space based on the results.
Then, a multiple-interference-channel based MPB is
proposed according to the principles. In order to adapt to dynamic
channels, Section IV derives an adaptive algorithm for the proposed
beamformer. Finally, Section V gives a number of computation and
simulation results that illustrate the good performance of this
beamformer, and Section VI concludes the paper.

\section{Problem Formulation}
\label{Sec:ProblemForm}

\subsection{Signal Model}
\label{Sec:ProblemForm:SigModel}

In a CDMA system with $M$ users, the transmitted baseband signal of
the $i$th user is
\begin{equation}
\label{Equ:Signal_model:tranmit_baseband_User_i} s_i(t) =
\sqrt{P_T}\sum_{k=-\infty}^{+\infty} b_i(k) c_i(t-kT_s)
\end{equation}
where $P_T$ is the transmit power; $b_i(k) \in \{+1,\,-1\}$ is the
$k$th transmitted symbol by the $i$th user; $c_i(t)$ is its
normalized signaling waveform, supported on $[0,\,T_s]$; and $T_s$
denotes the symbol interval. $c_i(t)$ can be expressed as
\begin{equation}
c_i(t) = \sum_{n=0}^{N-1} C_i(n) \psi(t-nT_c)
\end{equation}
where $C_i(n)\in\{+1,-1\}$ is the spreading code assigned to the
$i$th user; $\psi(t)$ is the normalized chip waveform with time
duration $T_c$; and $N = T_s/T_c$ is the processing gain.

The receiver has an antenna array of $L$ isotropic elements that
receives signals from far field. Each user signal arrives at the
array via different paths. We assume all elements experience
identical fading for each path. In addition, there are $Q$ jammings
received. Then the total received signal after carrier demodulation
is
\begin{equation}
\label{Equ:Signal_model:received_analog} \mathbf{x}(t) \!=\!
\sum_{i=0}^{M-1}\! \sum_{j=1}^{D_i} \! \alpha_{ij}
s_i(t\!-\!\tau_{ij}) \mathbf{a}(\theta_{ij}) \!+\! \sum_{q=1}^{Q} \!
z_q(t) \mathbf{a}(\theta_q)  \!+\!  \mathbf{v}(t)
\end{equation}
where $\alpha_{ij}$, $\tau_{ij}$ and $\mathbf{a}(\theta_{ij})$ are
the path gain, delay and array response vector for the $j$th path of
the $i$th user; $D_i$ is the number of paths for the $i$th user;
$z_q(t)$ and $\mathbf{a}(\theta_{q})$ are the waveform and the array
response vector for the $q$th jamming; $\mathbf{v}(t)$ is the
space-time white noise. For uniform linear array (ULA) with
interelement spacing $d$ and carrier wavelength $\lambda$, the $l$th
component of $\mathbf{a}(\theta)$ is $ e^{-j \frac{2 \pi l
d}{\lambda} \sin (\theta)}$, where $\theta$ is the DOA and can be
$\theta_q$ or $\theta_{ij}$.

After matched filtering and chip-rate sampling, the discrete signal
can be written as
\begin{align}
\label{Equ:Signal_model:received_discrete} \mathbf{x}(n) &=
\int_{nT_c}^{(n+1)T_c} \mathbf{x}(t) \psi^{\ast}(t-nT_c) dt
\nonumber\\
&=\sum_{i=0}^{M-1}\sum_{j=1}^{D_i}\sqrt{P_{ij}}
\sum_{k=-\infty}^{+\infty} b_i(k)c_i(n-n_{ij}-kN)
\mathbf{a}(\theta_{ij}) + \sum_{q=1}^{Q} z_q(n) \mathbf{a}(\theta_q) +\mathbf{v}(n)
\end{align}
where $(\cdot)^\ast$ denotes conjugate; $P_{ij}$ and $n_{ij}$ are power and chip delay for the $j$th
path of the $i$th user, respectively. We have omitted $\alpha_{ij}$
in (\ref{Equ:Signal_model:received_discrete}) and contained it in
$P_{ij}$; $z_q(n)$ and $\mathbf{v}(n)$ are the discrete counterpart
of $z_q(t)$ and $\mathbf{v}(t)$.

We also assume the propagation delays of multipath signals from a
desired user, enumerated as $i=0$ in
(\ref{Equ:Signal_model:received_discrete}), can be perfectly
estimated as the existing MPBs\cite{Naguib1996,Song2001,choi2002nab,Yang2006fbba,torrieri2004,torrieri2007maa},
and our goal is to recover $b_0(k)$ from $\mathbf{x}(n)$ with
fidelity. There are $D_0$ paths for the desired user, and our
strategy is to construct beamformer for each path to suppress all
other signals except the specified path. In fact, the delayed
replica of the desired signal in the multipath propagation can be
treated as MAIs when the relative delay between a certain path and
the desired one is greater than one chip, since the spreading code
is assumed to have good cross-correlation and self-correlation
property. Then, a two-dimensional rake combiner is employed to
combine outputs of the $D_0$ beamformers, and the procedure is
similar to \cite{choi2002nab,Yang2006fbba}. Since the main purpose
of this paper is to address the problem of the threshold effect of
the MPB, without loss of generality, the first beamformer
(corresponding to the first path of the desired user) is used for
the following analysis for notational convenience. To be more
specific, we rewrite (\ref{Equ:Signal_model:received_discrete}) as
    \begin{equation}
    \label{Equ:Signal_model:x} \mathbf{x}(n)= \sum_{i=0}^{D}\sqrt{P_i}
    s_i(n) \mathbf{a}(\theta_i) + \mathbf{v}(n),
    \end{equation}
where $D=\sum_{i=0}^{M-1}D_{i}+Q-1 < L$; $s_i(n)$ is the discrete
sequence of the $i$th signal with normalized power, with $s_0(n)$ is
the SOI, and $s_1(n),s_2(n),\ldots,s_D(n)$ are interferers such as
other multipath signals of the desired user, MAIs by other $M-1$
users, and jammers, etc. $P_i$, $\mathbf{a}(\theta_i)$, and
$\theta_i$ are its power, steering vector and DOA, respectively.
Specifically, the SOI $s_0(n)$ is

    \begin{equation}
    \label{Equ:Signal_model:s0}
    s_0(n)=\sum_{k=-\infty}^{+\infty}b_0(k)c_0(n-kN-n_0),
    \end{equation}
where $n_0=n_{01}$ is the equivalent propagation delay.

\subsection{The Matrix Pair Beamformer}
\label{Sec:ProblemForm:MPBeamformer}

The steering vector $\mathbf{a}(\theta_i)$ in
(\ref{Equ:Signal_model:x}) is a spatial signature of the $i$th
signal, which is different from others so long as they arrive from
different directions. Beamformer is a spatial filter that exploits
such difference to pass the desired signal $s_0(n)$ while
suppressing $s_1(n)\ldots s_D(n)$ and $\mathbf{v}(n)$. A
statistically optimum beamformer \cite{Trees2002} generally requires
at least, either explicitly or implicitly, the information about the
steering vector $\mathbf{a}(\theta_0)$ and the interference
covariance matrix. The latter one may be replaced by the data
covariance matrix, so the remaining problem is how to acquire
$\mathbf{a}(\theta_0)$. To work ``blindly'', i.e. without explicit
information of DOA, the methods in
\cite{Naguib1996,Song2001,choi2002nab,Yang2006fbba,torrieri2004,torrieri2007maa}
exploit the temporal signature of the desired signal to acquire
these spatial statistical information. Specifically, it is
implemented by two orthogonal projection operations and a
generalized eigen-decomposition to exploit a
``\emph{mismatch--match}'' mechanism in a covariance matrix pair.
Hence, we refer to them as matrix pair beamformer
\cite{Jianshu,Jianshu2}. With the data segmentation, the array
outputs corresponding to the $k$th symbol of the SOI can be
expressed in the following matrix form:
    \begin{align}
    \label{Equ:MPBeamformer:X}
    \mathbf{X}(k) \! &\triangleq \! \Big[ \; \mathbf{x}(kN+n_0) \; \cdots \; \mathbf{x}(kN+n_0+N-1) \; \Big]
    \nonumber\\
                     &=          \! \left[\sqrt{P_0} b_0(k)\right] \! \mathbf{a}_0\mathbf{c}_0^T
                                 \!\!+\!\! \sum_{i=1}^{D}\sqrt{P_i}\mathbf{a}_i \mathbf{s}_i^T(k)
                                 \!\!+\!\! \mathbf{V}(k)
                                 \nonumber\\
                     &=          \! \left[\sqrt{P_0} b_0(k)\right] \! \mathbf{a}_0\mathbf{c}_0^T
                                 \!\!+\!\! \mathbf{A}_I \pmb{\Theta}_I^{\frac{1}{2}} \mathbf{S}_I^T(k)
                                 \!\!+\!\! \mathbf{V}(k),
    \end{align}
where $\mathbf{a}_i$ stands for $\mathbf{a}(\theta_i)$,
($i\in\{0,1,\ldots,D\}$) and $\mathbf{A}_I$ is a matrix whose
columns are the steering vectors of interferers $\mathbf{a}_1\ldots
\mathbf{a}_D$; $\mathbf{c}_0$ is the temporal signature vector of
the SOI composed of the spreading code and $(\cdot)^{T}$ denotes
transpose; $\mathbf{V}(k)$ are the matrix form of the noise;
$\mathbf{s}_i(k)$ are the matrix form of the $i$th interferer and
$\mathbf{S}_I(k)$ is the matrix whose columns are
$\mathbf{s}_1(k)\ldots \mathbf{s}_D(k)$, with
    \begin{align}
    \mathbf{A}_I &\triangleq \big[ \; \mathbf{a}_1 \; \mathbf{a}_2 \; \cdots \; \mathbf{a}_D \; \big]
    \nonumber\\
    \mathbf{c}_0 &\triangleq \big[ \; c_0(0) \; c_0(1) \; \cdots \; c_0(N-1)
    \; \big]^T
    \nonumber\\
    \mathbf{s}_i(k) &\triangleq \big[ \;
    s_i(kN+n_0) \; \cdots \; s_i(kN+n_0+N-1) \; \big]^T
    \nonumber\\
    \mathbf{S}_I(k) &\triangleq \big[ \;
    \mathbf{s}_1(k) \; \mathbf{s}_2(k) \; \cdots \; \mathbf{s}_D(k) \; \big]
    \nonumber\\
    \mathbf{V}(k) &\triangleq \big[ \;
    \mathbf{v}(kN+n_0) \; \cdots \; \mathbf{v}(kN+n_0+N-1) \; \big]
    \nonumber\\
    \pmb{\Theta}_I  &\triangleq \mathrm{diag}\big\{P_1,P_2,\cdots,P_D\big\}.    \nonumber
    \end{align}

Then, the $k$th data block in each antenna is projected onto two
subspaces: signal space $\mathcal{S}$ and interference space
$\mathcal{I}$, respectively. $\mathcal{S}$ is a one-dimensional
space with base vector
$\mathbf{h}_\mathcal{S}=\mathbf{c}_0/\sqrt{N}$, and $\mathcal{I}$ is
a specifically designed $r_\mathcal{I}$-dimensional space with base
vectors
$\mathbf{h}_\mathcal{I}^{(1)},\ldots,\mathbf{h}_\mathcal{I}^{(r_\mathcal{I})}$.
The projection operation produces signal snapshot
$\mathbf{x}_\mathcal{S}(k)$ and the interference snapshot
$\mathbf{X}_\mathcal{I}(k)$. Define $\mathbf{H}_\mathcal{I}
\triangleq \big[ \mathbf{h}_\mathcal{I}^{(1)} \;
\mathbf{h}_\mathcal{I}^{(2)} \; \cdots \;
\mathbf{h}_\mathcal{I}^{(r_\mathcal{I})} \big]$ and assume
$\mathbf{H}_\mathcal{I}^H\mathbf{H}_\mathcal{I} = \mathbf{I}$, where
$(\cdot)^H$ denote conjugate and transpose. Then the projection procedures may be
written as
    \begin{align}
    \label{Equ:MPBeamformer:x_s}
    &\mathbf{x}_\mathcal{S}(k)
                    =  \mathbf{X}(k)\mathbf{h}_S^\ast       \nonumber\\
                    &=  \! \left[\sqrt{NP_0} b_0(k)\right]  \mathbf{a}_0
                        \!+\! \frac{1}{\sqrt{N}}\mathbf{A}_I \pmb{\Theta}_I^{\frac{1}{2}} \mathbf{S}_I^T(k)\mathbf{c}_0^\ast
                        \!+\! \mathbf{v}_\mathcal{S}(k)\\
    \label{Equ:MPBeamformer:x_I}
    &\mathbf{X}_\mathcal{I}(k)
                    =  \mathbf{X}(k)\mathbf{H}_\mathcal{I}^\ast      \nonumber\\
                    &=  \! \left[ \! \sqrt{P_0} b_0(k) \! \right] \mathbf{a}_0\mathbf{c}_0^T\mathbf{H}_\mathcal{I}^\ast
                        \!+\! \mathbf{A}_I \pmb{\Theta}_I^{\frac{1}{2}} \mathbf{S}_I^T(k)\mathbf{H}_\mathcal{I}^\ast
                        \!+\! \mathbf{V}_\mathcal{I}(k),
    \end{align}
where $\mathbf{v}_\mathcal{S}(k) =
\mathbf{V}(k)\mathbf{h}_\mathcal{S}^\ast$ and
$\mathbf{V}_\mathcal{I}(k) =
\mathbf{V}(k)\mathbf{H}_\mathcal{I}^\ast$.

Assume the SOI is uncorrelated with the interferers, we can derive
the covariance matrices of $\mathbf{x}_\mathcal{S}(k)$ and
$\mathbf{X}_\mathcal{I}(k)$ as
    \begin{align}
        \label{Equ:MPBeamformer:R_S}
        \mathbf{R}_\mathcal{S} &\triangleq E\big\{\mathbf{x}_\mathcal{S}(k) \mathbf{x}_\mathcal{S}^H(k)\big\} = \sigma_{\mathcal{S}_0}^2 \mathbf{a}_0 \mathbf{a}^H_0 + \mathbf{Q}_\mathcal{S}        \\
        \label{Equ:MPBeamformer:R_I}
        \mathbf{R}_\mathcal{I} &\triangleq \frac{1}{r_\mathcal{I}}E\big\{\mathbf{X}_\mathcal{I}(k) \mathbf{X}_\mathcal{I}^H(k)\big\} = \sigma_{\mathcal{I}_0}^2 \mathbf{a}_0 \mathbf{a}^H_0 +\mathbf{Q}_\mathcal{I},
    \end{align}
where
    \begin{align}
        \sigma_{S_0}^2              &=      P_0 \mathbf{c}_0^H\mathbf{P}_\mathcal{S}\mathbf{c}_0 = NP_0   \label{Equ:MPBeamformer:sigma_S0}\\
        \sigma_{\mathcal{I}_0}^2    &=      \frac{P_0}{r_\mathcal{I}}\mathbf{c}_0^H\mathbf{P}_\mathcal{I}\mathbf{c}_0,     \label{Equ:MPBeamformer:sigma_I0}
    \end{align}
$\mathbf{Q}_\mathcal{S}$ and $\mathbf{Q}_\mathcal{I}$ are the
covariance matrices of the last two terms in
(\ref{Equ:MPBeamformer:x_s}) and (\ref{Equ:MPBeamformer:x_I}),
respectively. $\mathbf{P}_\mathcal{S}$ and $\mathbf{P}_\mathcal{I}$
are the projection matrices of $\mathcal{S}$ and $\mathcal{I}$,
defined as
    \begin{align}
        \mathbf{P}_\mathcal{S}      &=      \mathbf{h}_\mathcal{S}\mathbf{h}_\mathcal{S}^H = \frac{1}{N} \mathbf{c}_0\mathbf{c}_0^H    \label{Equ:MPBeamformer:P_S} \\
        \mathbf{P}_\mathcal{I}      &=      \mathbf{H}_\mathcal{I}\mathbf{H}_\mathcal{I}^H = \sum_{r=1}^{r_\mathcal{I}} \mathbf{h}_\mathcal{I}^{(r)}[\mathbf{h}_\mathcal{I}^{(r)}]^H.     \label{Equ:MPBeamformer:P_I}
    \end{align}
In practice, $\mathbf{R}_\mathcal{S}$ and $\mathbf{R}_\mathcal{I}$
are computed by sample averaging (c.f. Section \ref{Sec:MIC-MPB:RecursiveAlgorithm}). 

In most of the existing approaches, $\mathcal{I}$ is one dimensional
space ($r_\mathcal{I}=1$). The pre- and post-correlation (PAPC)
scheme\cite{Naguib1996,choi2002nab,Yang2006fbba,Song2001} uses
$\mathbf{x}(n)$ to calculate $\mathbf{R}_\mathcal{I}$, thus it is
equivalent to selecting one column of $\mathbf{I}_{N \times N}$ as
$\mathbf{H}_\mathcal{I}$, i.e.
    \begin{equation}
    \label{Equ:MPBeamformer:h_I_PAPC}
    \mathbf{H}_\mathcal{I}
    =
    \big[ \; 0 \;\; \cdots \;\; 0 \;\; 1 \;\; 0 \;\; \cdots \;\; 0 \;\big]^T.
    \end{equation}
The Maximin scheme in \cite{torrieri2004} and \cite{torrieri2007maa}
employs a monitor filter to isolate the interference, which can be
interpreted as
    \begin{equation}
    \label{Equ:MPBeamformer:h_I_FP}
    \mathbf{H}_\mathcal{I}
    =
    \mathbf{c}_0 \odot \left[ \; 1 \;\; e^{j2 \pi f_\mathrm{MF}} \;\; \cdots \;\;
    e^{j2\pi f_\mathrm{MF} (N-1)} \;\right]^T,
    \end{equation}
where $f_\mathrm{MF}\in(0,1]$ is the normalized center frequency of
the monitor filter (MF), and $\odot$ denotes the Hadamard product.

Under the maximum signal-to-interference-plus-noise ratio (MSINR)
criterion, it is well known that the optimal weight vector for the
first propagation path of the desired user $\mathbf{w}_\mathrm{opt}$
is the generalized eigenvector corresponding to the largest
generalized eigenvalue of the matrix pair
$(\mathbf{R}_\mathcal{S},\mathbf{R}_\mathcal{I})$, i.e.,
    \begin{equation}
    \label{Equ:MPBeamformer:EigenEquation}
    \mathbf{R}_\mathcal{S}\mathbf{w}_\mathrm{opt} = \lambda_\mathrm{max}
    \mathbf{R}_\mathcal{I}\mathbf{w}_\mathrm{opt},
    \end{equation}
where $\lambda_\mathrm{max}$ is the largest generalized eigenvalue.
Therefore, the MPB can maximize the output
signal-to-interference-plus-noise ratio (SINR) when
$\mathbf{w}_\mathrm{opt}$ is applied to $\mathbf{x}_\mathcal{S}(k)$,
and the output $y_{o}(k)$ is
    \begin{equation}
    \label{Equ:MPBeamformer:y_out}
    y_{o}(k)=\mathbf{w}_\mathrm{opt}^{H}\mathbf{x}_\mathcal{S}(k)=y_{S}(k)+y_{I}(k)+y_{N}(k),
    \end{equation}
where
\begin{align}
        y_{S}(k)              &=      \big[\sqrt{NP_{0}}b_{0}(k)\big]\mathbf{w}_\mathrm{opt}^{H}\mathbf{a}_0        \nonumber\\
        y_{I}(k)              &=      \frac{1}{\sqrt{N}}\mathbf{w}_\mathrm{opt}^{H}\mathbf{A}_I \pmb{\Theta}_I^{\frac{1}{2}} \mathbf{S}_I^T(k)\mathbf{c}_0^\ast       \nonumber\\
        y_{N}(k)              &=      \mathbf{w}_\mathrm{opt}^{H}\mathbf{v}_{\mathcal{S}}(k).   \nonumber
    \end{align}
Then, the final array output after a two-dimensional rake combiner
can be written as \cite{choi2002nab,Yang2006fbba}
\begin{equation}
    z(k)=\sum_{j=1}^{D_{0}}y_{j,o}(k),
\end{equation}
where $y_{j,o}(k)$ is the $j$th output of the beamformer
corresponding to the $j$th propagation path, and the typical
expression of $y_{j,o}(k)$ can be referred to
(\ref{Equ:MPBeamformer:y_out}).

\subsection{Threshold Effects Regarding MPB}
\label{Sec:ProblemForm:MPResults} Based on the theoretical analysis
in \cite{Jianshu,Jianshu2}, $\lambda_\mathrm{max}$ has the following
property:
    \begin{equation}
    \label{Equ:MPBeamformer:LambdaMax}
    \lambda_\mathrm{max} \approx \mathrm{max}\big\{\gamma_{0}+1,
    \gamma_{1}+1\big\},
    \end{equation}
where
    \begin{equation}
    \label{Equ:MPBeamformer:Gamma0}
    \gamma_{0}=\frac{L(N-\beta)\textsf{SNR}}{L\beta\textsf{SNR}+N}
    \end{equation}
is a monotonically increasing function of $\textsf{SNR}$, and
$\textsf{SNR} \triangleq \sigma_{\mathcal{S}_0}^2/\sigma^2$
        is the SNR of the SOI after despreading (or equivalently, input SNR per symbol). $\beta$ is the normalized power leakage ratio (PLR) in interference channel defined as
    \begin{align}
    \label{Equ:MPBeamformer:beta}
        \beta           &\triangleq         \frac{\sigma_{\mathcal{I}_0}^2}{P_0}
                        =                   N\frac{\sigma_{\mathcal{I}_0}^2}{\sigma_{\mathcal{S}_0}^2}
                        =
                        \frac{\mathbf{c}_0^H\mathbf{P}_\mathcal{I}\mathbf{c}_0}{r_\mathcal{I}};
    \end{align}
$\gamma_{1}+1$ is the the largest generalized eigenvalue of the
matrix pair $(\mathbf{Q}_\mathcal{S},\mathbf{Q}_\mathcal{I})$, which
is co-determined by the structure and power of interferers as well
as the projection spaces of the MPB. It can
be derived that $\gamma_{1}$ could be bounded if the following
expression is satisfied \cite{Jianshu,Jianshu2}
\begin{equation}
\label{Equ:MPBeamformer:SpaceCondition} \mathcal{I}^{\perp}\cap
\mathcal{V}_{I}\subseteq \mathcal{S}^{\perp}\cap \mathcal{V}_{I},
\end{equation}
where
    \begin{align}
    \mathcal{I} & \triangleq \mathcal{R}\big\{\mathbf{P}_{\mathcal{I}}\big\} \nonumber\\
    \mathcal{S} & \triangleq \mathcal{R}\big\{\mathbf{P}_{\mathcal{S}}\big\}\nonumber\\
    \mathcal{V}_{I} & \triangleq \mathrm{span}\big\{\mathbf{S}_I(1)\big\} ,\nonumber
    \end{align}
where $(\cdot)^{\perp}$ denotes orthogonal complement space,
$\mathcal{R}(\cdot)$ denotes the range space of a matrix;
$\mathcal{V}_{I}$ is the space spanned by interference sequences and
$\mathbf{S}_{I}(1)$ are the waveforms of the interferers in the
first period.

The optimal weight vector $\mathbf{w}_\mathrm{opt}$ can be approximated by the following
equation \cite{Jianshu,Jianshu2}

\begin{equation}
\label{Equ:MPBeamformer:wopt} \mathbf{w}_\mathrm{opt} \approx
\begin{cases}
\mu_{1}\cdot\mathbf{R}_\mathcal{I}^{-1}\mathbf{a}_0& \text{if $\gamma_{0}>\gamma_{1}$}\\
\mu_{2}\cdot\mathbf{R}_\mathcal{I}^{-1}\mathbf{a}_{\epsilon_1}&
\text{if $\gamma_{0}<\gamma_{1}$},
\end{cases}
\end{equation}
where $\mathbf{a}_{\epsilon_1}$ is an appropriate linear combination
of the steering vectors of interferers
$\mathbf{a}_1,\mathbf{a}_2,\ldots,\mathbf{a}_D$, and $\mu_{1},\,
\mu_{2} $ are the coefficients. The expression of
$\mathbf{w}_\mathrm{opt}$ means that if $\gamma_{0}>\gamma_{1}$, the
main beam of the MPB will point to the DOA of the SOI; if
$\gamma_{0}<\gamma_{1}$, the main beam of the MPB will
point to the DOA of the interferers. Furthermore, if $\beta\neq 0$,
the beamformer will form a notch in the direction of the SOI because
$\mathbf{R}_\mathcal{I}$ contains parts of the desired signal.

Our work also shows that the existing MPBs are vulnerable
to structured interference, such as periodically repeated white
noise, tones, and MAIs in multipath channels for
(\ref{Equ:MPBeamformer:SpaceCondition}) can hardly be satisfied in
some cases of those scenarios. For periodical interference,
(\ref{Equ:MPBeamformer:SpaceCondition}) can be rewritten as the
following \cite{Jianshu,Jianshu2}

\begin{equation}
\label{Equ:MPBeamformer:SpaceCondition2}
\mathcal{R}\big\{\mathbf{P}_{\mathcal{V}_{I}}\,
\mathbf{H}_{\mathcal{I}}\big\} \supseteq
\mathcal{R}\big\{\mathbf{P}_{\mathcal{V}_{I}}\,
\mathbf{h}_{\mathcal{S}}\big\},
\end{equation}
or equivalently,
\begin{equation}
\label{Equ:MPBeamformer:SpaceCondition2b} \mathcal{R}\big\{
\mathbf{H}_{\mathcal{V}_{I}} \mathbf{H}_{\mathcal{V}_{I}}^H
\mathbf{H}_\mathcal{I}\big \} \supseteq \mathcal{R}\big\{
\mathbf{H}_{\mathcal{V}_{I}} \mathbf{H}_{\mathcal{V}_{I}}^H
\mathbf{h}_\mathcal{S} \big\}.
\end{equation}
where $\mathbf{P}_{\mathcal{V}_{I}}$ is the projection matrix of the
subspace $\mathcal{V}_{I}$, $\mathbf{H}_{\mathcal{V}_{I}}$ is a base
matrix of the subspace
$\mathcal{V}_{I}\triangleq\mathcal{R}\big\{\mathbf{H}_{\mathcal{V}_{I}}\big\}=\mathcal{R}\big\{\mathbf{S}_{I}(1)\big\}$.
If (\ref{Equ:MPBeamformer:SpaceCondition2}) does not hold,
$\gamma_{1}+1$ will grow as the interference power increases.

\begin{figure*}[!t]
\centering
\includegraphics[width=0.8\textwidth]{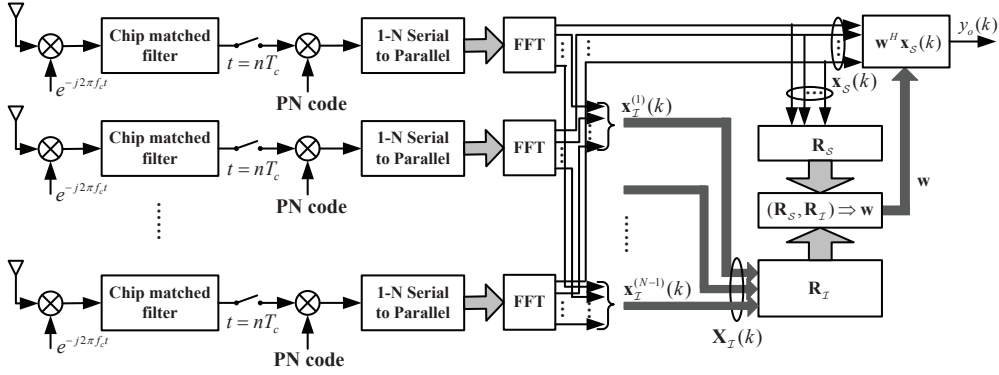}
\caption{Projection operations to separate the signal channel and
the interference channels using FFT base vectors.}
\label{Fig:MIC-MPB:Proj-DFT}
\end{figure*}

From the above discussion, we see that the threshold effects of MPBs rely heavily on base matrix $\mathbf{H}_\mathcal{I}$ for
the interference space $\mathcal{I}$. Therefore, in the following
section, we will propose appropriate methods to handle this effect
as well as finite sample performance by designing appropriate base
vectors for the interference space.

\section{The Multiple Interference Channel based MPB}
\label{Sec:MIC-MPB} 
In this section, starting from the above results, we first present
the principles for designing projection space for MPBs.

\subsection{Principles for Designing Projection Space for MPBs}
\label{Sec:MIC-MPB:ProjectionDesign}

\subsubsection{Ability of Suppressing Structured Interference}
\label{Sec:MIC-MPB:ProjectionDesign:ASPI} Since an MPB can
work properly only if $\gamma_{0}>\gamma_{1}$, $\gamma_{0}$ should
be as large as possible for a given $\textsf{SNR}$.
(\ref{Equ:MPBeamformer:Gamma0}) shows that $\gamma_{0}$ is a
monotonically decreasing function of $\beta$, so $\beta$ should be
designed as small as possible. It can also be found from
(\ref{Equ:MPBeamformer:R_I}) and (\ref{Equ:MPBeamformer:wopt}) that,
if $\beta \neq 0$, there will be the sample-correlation terms
between the SOI and the interference-plus-noise in
$\mathbf{R}_\mathcal{I}$ because of finite sample effects. Even if
$\gamma_{0}>\gamma_{1}$, the sample-correlation terms will cause the
main-lobe unstable as well as a ¡°signal cancellation¡± effect in
the beamformer output \cite{WaxApr1996,Wax2Apr1996}. Therefore,
$\beta$ should be designed to be $0$. With
(\ref{Equ:MPBeamformer:beta}), we can easily derive that
\begin{equation}\label{Equ:MIC-MPB:principle1}
\beta = 0\,\,\,\,\,\Leftrightarrow\,\,\,\,\,\mathcal{I} \subseteq
\mathcal{S}^{\perp}.
\end{equation}
On the other hand, $\gamma_{1}$ should be as small as possible for
given power of interference.
(\ref{Equ:MPBeamformer:SpaceCondition2}) means the subspace spanned
by the columns of $\mathbf{H}_{\mathcal{I}}$ projected onto
$\mathcal{V}_{I}$ must contain the subspace spanned by
$\mathbf{h}_{\mathcal{S}}$ projected onto $\mathcal{V}_{I}$. Since
$\mathcal{R}\big\{\mathbf{H}_{\mathcal{V}_{I}}
\mathbf{H}_{\mathcal{V}_{I}}^H \mathbf{h}_\mathcal{S}
\big\}\subseteq \mathcal{V}_{I}$,
(\ref{Equ:MPBeamformer:SpaceCondition2}) always holds so long as
$\mathcal{R}\big\{\mathbf{H}_{\mathcal{V}_{I}}
\mathbf{H}_{\mathcal{V}_{I}}^H \mathbf{H}_\mathcal{I} \big\}=
\mathcal{V}_{I}$, which means the columns of
$\mathbf{H}_\mathcal{I}^{H}\mathbf{H}_{\mathcal{V}_{I}}$ are linear
independent, i.e.,
\begin{equation}\label{Equ:MIC-MPB:principle2}
\forall\, \pmb{\eta} \neq \mathbf{0},\,\,\,\,
\mathbf{H}_\mathcal{I}^H\mathbf{H}_{\mathcal{V}_{I}} \cdot
\pmb{\eta} \neq \mathbf{0}.
\end{equation}
This expression shows the subspace $\mathcal{I}$ should be properly
designed in order that the subspace
$\mathcal{V}_{I}=\mathcal{R}\big\{\mathbf{S}_{I}(1)\big\}$ does not
contain any vector which is perpendicular to the subspace
$\mathcal{R}\big\{\mathbf{H}_\mathcal{I}\big\}=\mathcal{I}$.

\subsubsection{Improving Finite Sample Size Performance}
\label{Sec:ProblemForm:ProjectionDesign:IFSSP}

If $\beta = 0$ and $\mathbf{R}_\mathcal{I}$ does not contain any
component of the SOI, the beamformer can be considered as an Miminum
Variance Distortionless Response (MVDR) beamformer when
$\gamma_0>\gamma_1$ by (\ref{Equ:MPBeamformer:wopt}), and the
performance of the beamformer is degraded mostly by the disturbed
noise space \cite{Trees2002} and at least $K\approx 2L$ samples of
data are needed to maintain an average loss ratio of better than
one-half (less than 3 dB) \cite{Reed1973}. It can be considered that
the number of independent noise samples available is the number of
the effective samples. We now examine the relationship between the
number of effective samples and $\mathcal{I}$. From
(\ref{Equ:MPBeamformer:x_I}), the $r$th column of
$\mathbf{X}_\mathcal{I}(k)$ can be written as
\begin{align}
\label{Equ:MPBeamformer:x_I_r}
        \mathbf{x}_\mathcal{I}^{(r)}(k)     &=      \big[\sqrt{P_0} b_0(k)\big] \mathbf{a}_0\mathbf{c}_0^T\big[\mathbf{h}_\mathcal{I}^{(r)}\big]^\ast
                                                    + \mathbf{A}_I \pmb{\Theta}_I^{\frac{1}{2}}
                                                    \mathbf{S}_I^T(k)\big[\mathbf{h}_\mathcal{I}^{(r)}\big]^\ast + \mathbf{v}_{\mathcal{I},r}(k), \,\,\,\,r =
                                            1,2,\ldots,r_\mathcal{I}
\end{align}
where $\mathbf{v}_{\mathcal{I},r}(k) \triangleq
\mathbf{V}(k)[\mathbf{h}_\mathcal{I}^{(r)}]^\ast$. Since all
elements of $\mathbf{V}(k)$ are \textit{i.i.d} zero-mean Gaussian
random variables, it can be easily obtained that
\begin{equation}
        E\left\{ \mathbf{v}_{\mathcal{I},r}(k) \mathbf{v}_{\mathcal{I},r'}^H(k') \right\} = \sigma^2 \delta_{rr'} \delta_{kk'}
        \mathbf{I},
\end{equation}
i.e., the noise component $\mathbf{v}_{\mathcal{I},r}(k)$ of
different $\mathbf{x}_\mathcal{I}^{(r)}(k)$ is mutually independent.
As a result, the number of the effective samples extracted per data
symbol is $r_\mathcal{I}$, and the total number of the effective
samples is $K\cdot r_\mathcal{I}$ with $K$ symbols. This result
shows that the dimension $r_\mathcal{I}$ of subspace $\mathcal{I}$
determines the finite sample performance of an MPB.

\subsection{The Multiple Interference Channel based MPB}
\label{Sec:MIC-MPB} According to (\ref{Equ:MIC-MPB:principle1}) and
(\ref{Equ:MIC-MPB:principle2}), we can select the subspace
$\mathcal{I}$ as the following equation
\begin{equation}\label{Equ:MIC-MPB:I}
\mathcal{I}     =       \mathcal{S}^\perp          =
\mathrm{span}\{\mathbf{c}_0\}^\perp.
\end{equation}
Since only vectors in $\mathrm{span}\{\mathbf{c}_0\}$ can be
perpendicular to $\mathcal{I}$, there is no vector in
$\mathcal{R}\big\{ \mathbf{S}_I(1) \big\}$ which is perpendicular to
$\mathcal{I}$ so long as $\mathbf{c}_0 \notin
\mathcal{R}\big\{\mathbf{S}_I(1)\big\}$. This condition can be
easily satisfied in most cases in a multi-user CDMA system. On the
other hand, the dimension $r_\mathcal{I}$ of the subspace
$\mathcal{I}$ equals to $N-1$ under this condition, then the
effective number of samples obtained per symbol is also $N-1$, which
is the maximum value obtained when $\beta=0$.

Specifically, we select the following vector as the the $r$th
($r=1,\ldots,N-1$) base vector of the subspace $\mathcal{I}$,
\begin{equation}
        \label{Equ:MIC-MPB:h_I_r}
        \mathbf{h}_{\mathcal{I},\mathrm{MIC}}^{(r)}    =   \frac{1}{\sqrt{N}}\mathbf{c}_0 \odot
        \mathbf{W}_N^r,
\end{equation}
where
$\big\{\mathbf{W}_N^0,\mathbf{W}_N^1,\ldots,\mathbf{W}_N^{N-1}\big\}$
are the base vectors of the Discrete Fourier Transform (DFT), defined
as,
\begin{equation}
        \label{Equ:MIC-MPB:DFT_basis}
        \mathbf{W}_N^r  =   \left[
                                \begin{array}{cccc}
                                    1 & e^{j2\pi\frac{r}{N}} & \cdots & e^{j2\pi\frac{r(N-1)}{N}}
                                \end{array}
                            \right]^T.
\end{equation}

Comparing with the Maximin or PAPC method which has only one vector
in interference channel (or equivalently, subspace $\mathcal{I}$ ),
this method has $N-1$ base vectors, so it can be called
Multiple-Interference-Channel Matrix Pair Beamformer (MIC-MPB).

If we define an $L\times N$ matrix
\begin{equation}
\mathbf{C}_0=[\mathbf{c}_{0}\,\,\mathbf{c}_{0}\,\,\cdots\,\,\mathbf{c}_{0}]^{H}/\sqrt{N}\nonumber,
\end{equation}
an $L\times N$ matrix
\begin{equation}
\mathbf{X}_H(k)=[\mathbf{x}_\mathcal{S}(k)\,\,\mathbf{x}_\mathcal{I}^{(1)}(k)\,\,\cdots\,\,\mathbf{x}_\mathcal{I}^{(N-1)}(k)]\nonumber,
\end{equation}
and an $N\times N$ matrix
\begin{equation}
\mathbf{W}=[\mathbf{W}_{N}^{0}\,\,\mathbf{W}_{N}^{1}\,\,\cdots\,\,\mathbf{W}_{N}^{N-1}]\nonumber,
\end{equation}
it can be easily obtained from (\ref{Equ:MPBeamformer:x_s}) and
(\ref{Equ:MPBeamformer:x_I_r})
\begin{align}
\label{Equ:MIC-MPB:X_H_k}
\mathbf{X}_H(k)     &=      \mathbf{X}(k)
\left[
                                                            \begin{array}{cccc}
                                                                \frac{1}{\sqrt{N}}\mathbf{c}_{0} & \mathbf{h}_{\mathcal{I},\mathrm{MIC}}^{(1)} & \cdots & \mathbf{h}_{\mathcal{I},\mathrm{MIC}}^{(N-1)}
                                                            \end{array}
                                                    \right]^*\nonumber\\
                    &=      \big[\mathbf{X}(k) \odot \mathbf{C}_0\big]
                    \mathbf{W}^{*}.
    \end{align}

(\ref{Equ:MIC-MPB:X_H_k}) indicates the projection operations
implemented by the base vectors defined in (\ref{Equ:MIC-MPB:h_I_r})
are equivalent to the procedure illustrated in Fig.
\ref{Fig:MIC-MPB:Proj-DFT}. The zero frequency outputs of all DFTs
generate $\mathbf{x}_\mathcal{S}(k)$, and all $r$th frequency
outputs form $\mathbf{x}_\mathcal{I}^{(r)}(k)$. Mixing with the
spreading code flattens the spectrum of the interference and noise,
making the power evenly distributed on all frequencies. Furthermore,
using the DFT base vectors for projection operations can be
efficiently implemented by Fast Fourier Transform (FFT).

\section{Adaptive algorithm}
\label{Sec:MIC-MPB:RecursiveAlgorithm} In this section, we derive a
blind adaptive algorithm for the proposed MIC-MPB for each signal
path of the desired user. 
In order to adapt to time-varying environment, we use the exponentially weighted sample correlation matrices $\mathbf{R}_{\mathcal{S}}(k)$ and
$\mathbf{R}_{\mathcal{I}}(k)$ instead of $\mathbf{R}_{\mathcal{S}}$ and $\mathbf{R}_{\mathcal{I}}$. Then, the recursive update
equation for the matrices can be written as
    \begin{align}
        \label{Equ:MIC-MPB:R_S_Update}
        \mathbf{R}_{\mathcal{S}}(k) &= \mu \mathbf{R}_{\mathcal{S}}(k-1) + \mathbf{x}_\mathcal{S}(k) \mathbf{x}_\mathcal{S}^H(k)
        \\
        \label{Equ:MIC-MPB:R_I_Update}
        \mathbf{R}_{\mathcal{I}}(k) &= \mu \mathbf{R}_{\mathcal{I}}(k-1) + \mathbf{R}_\mathcal{I}^\Delta(k)
    \end{align}
where
    \begin{align}
        \mathbf{R}_\mathcal{I}^\Delta(k)     \triangleq      \frac{1}{N-1}
                                                            \sum_{r=1}^{N-1}
                                                            \mathbf{x}_\mathcal{I}^{(r)}(k)
                                                            \big[\mathbf{x}_\mathcal{I}^{(r)}(k)\big]^H
                                                            \nonumber
    \end{align}
and $\mu$ is a positive constant less than $1$. Since the update
term in (\ref{Equ:MIC-MPB:R_I_Update}) is not rank one, we cannot
apply Woodbury equality \cite{Meyer2000,Golub1996} to compute its
inverse. To solve this problem, let
$\mathbf{\hat{x}}_\mathcal{I}^{(r)}(k) \triangleq
\mathbf{x}_\mathcal{I}^{(r)}(k)/\sqrt{N-1}$ and define
    \begin{align}
        \label{Equ:MIC-MPB:R_I_Detla_k_t}
        \mathbf{R}_\mathcal{I}^\Delta(k;t)  &\triangleq \sum_{r=1}^{t} \mathbf{\hat{x}}_\mathcal{I}^{(r)}(k) \big[\mathbf{\hat{x}}_\mathcal{I}^{(r)}(k)\big]^H
        \\
        \label{Equ:MIC-MPB:R_I_k_t}
        \mathbf{R}_{\mathcal{I}}(k;t)       &\triangleq \mu \mathbf{R}_{\mathcal{I}}(k-1) +
        \mathbf{R}_\mathcal{I}^\Delta(k;t).
    \end{align}
Then we have $\mathbf{R}_\mathcal{I}^\Delta(k) =
\mathbf{R}_\mathcal{I}^\Delta(k;N-1)$,
$\mathbf{R}_{\mathcal{I}}(k;N-1) = \mathbf{R}_{\mathcal{I}}(k+1;0) =
\mathbf{R}_{\mathcal{I}}(k)$, and
$\mathbf{R}_\mathcal{I}^\Delta(k;t) =
\mathbf{R}_\mathcal{I}^\Delta(k;t-1)+\mathbf{\hat{x}}_\mathcal{I}^{(t)}(k)[\mathbf{\hat{x}}_\mathcal{I}^{(t)}(k)]^H$.
As a result, the following recursive equation can be obtained,
    \begin{equation}
        \label{Equ:MIC-MPB:R_I_k_t_Update2}
        \mathbf{R}_{\mathcal{I}}(k;t)       =       \mu(t) \cdot \mathbf{R}_{\mathcal{I}}(k;t-1)
                                                    +
                                                    \mathbf{\hat{x}}_\mathcal{I}^{(t)}(k)
                                                    \big[\mathbf{\hat{x}}_\mathcal{I}^{(t)}(k)\big]^H
    \end{equation}
where $\mu(t)$ is defined as
    \begin{equation}
        \label{Equ:MIC-MPB:mu_t}
        \mu(t)      =       \left\{
                                    \begin{array}{ll}
                                        \mu & t = 1\\
                                        1 & 2 \le t \le N-1
                                    \end{array}
                            \right.
    \end{equation}
We then apply Woodbury equality to
(\ref{Equ:MIC-MPB:R_I_k_t_Update2}) and obtain
    \begin{align}
        \label{Equ:MIC-MPB:R_I_inv_Update1}
        &\mathbf{c}(k;t)     =      \frac{
                                            [\mu(t)]^{-1}\mathbf{P}(k;t-1)\mathbf{\hat{x}}_\mathcal{I}^{(t)}(k)
                                        }
                                        {
                                            1
                                            +
                                            [\mu(t)]^{-1}
                                            \big[\mathbf{\hat{x}}_\mathcal{I}^{(t)}(k)\big]^H
                                            \mathbf{P}(k;t-1)\mathbf{\hat{x}}_\mathcal{I}^{(t)}(k)
                                        }
                                    \\
        \label{Equ:MIC-MPB:R_I_inv_Update2}
        &\mathbf{P}(k;t)\!     =      [\mu(t)]^{-1}\!
                                    \left\{
                                            \mathbf{I}\!-\!\mathbf{c}(k;t)\big[\mathbf{\hat{x}}_\mathcal{I}^{(t)}(k)\big]^H
                                    \!\right\}\!
                                    \mathbf{P}(k;t-1)
    \end{align}
when $t=N-1$, the value of $\mathbf{P}(k;t)$ are assigned to
$\mathbf{P}(k) \triangleq \mathbf{R}^{-1}_{\mathcal{I}}(k)$ and
reinitialization is need as the following,
    \begin{align}
        \label{Equ:MIC-MPB:P_k_t_Property1}
        &\mathbf{P}(k)      =       \mathbf{P}(k;N-1)
                                    \\
        \label{Equ:MIC-MPB:P_k_t_Property2}
        &\mathbf{P}(k+1;0)  =       \mathbf{P}(k;N-1).
    \end{align}
In summary, (\ref{Equ:MIC-MPB:R_S_Update}),
(\ref{Equ:MIC-MPB:mu_t}), (\ref{Equ:MIC-MPB:R_I_inv_Update1}),
(\ref{Equ:MIC-MPB:R_I_inv_Update2}),
(\ref{Equ:MIC-MPB:P_k_t_Property1}), and
(\ref{Equ:MIC-MPB:P_k_t_Property2}) complete the update of
$\mathbf{R}_{\mathcal{I}}(k)$ and $\mathbf{P}(k) =
\mathbf{R}_{\mathcal{I}}^{-1}(k)$. Then we can update the weight
vector $\mathbf{w}$ by power iterations \cite{Golub1996}:
    \begin{equation}
        \label{Equ:MIC-MPB:WeightVectorUpdate}
        \mathbf{w}(k+1)     =       \mathbf{P}(k)\mathbf{R}_{\mathcal{S}}(k)
        \frac{\mathbf{w}(k)}{\left\|\mathbf{w}(k)\right\|}.
    \end{equation}
The details of the algorithm are shown in Algorithm
\ref{Algorithm:MIC-MPB}.
\begin{algorithm}
    \caption{MIC-MPB Beamforming Alogrithm} \label{Algorithm:MIC-MPB}
    \begin{algorithmic}
        \begin{scriptsize}
            \STATE $\mathbf{R}_{\mathcal{S}}(0)=\delta\mathbf{I}$ where $\delta$ is a small positive number
            \STATE $\mathbf{P}(0,0) = \mathbf{P}(0) = \delta^{-1}\mathbf{I}$
            \STATE $\mathbf{w}(0) =
                                    [
                                    \begin{array}{cccc}
                                    1 & 0 & \cdots & 0
                                    \end{array}
                                    ]^T
                                    $
            \\
            \FOR{$k=1,2\ldots$}
                \STATE $\mathbf{R}_{\mathcal{S}}(k) = \mu \mathbf{R}_{\mathcal{S}}(k-1) + \mathbf{x}_\mathcal{S}(k) \mathbf{x}_\mathcal{S}^H(k)$
                \\
                \FOR{$t=1,2,\ldots,N-1$}
                    \IF{$t=1$}
                        \STATE $\mu(t) = \mu$
                    \ELSE
                        \STATE $\mu(t) = 1$
                    \ENDIF
                    \\
                    \STATE $\mathbf{\hat{x}}_\mathcal{I}^{(t)}(k) = \mathbf{x}_\mathcal{I}^{(t)}(k)/\sqrt{N-1}$
                    \STATE
                        $\mathbf{c}(k;t)
                        =
                        \frac{
                                \displaystyle
                                [\mu(t)]^{-1}\mathbf{P}(k;t-1)\mathbf{\hat{x}}_\mathcal{I}^{(t)}(k)
                            }
                            {
                                \displaystyle
                                1
                                +
                                [\mu(t)]^{-1}
                                [\mathbf{\hat{x}}_\mathcal{I}^{(t)}(k)]^H
                                \mathbf{P}(k;t-1)\mathbf{\hat{x}}_\mathcal{I}^{(t)}(k)
                            }$
                    \STATE
                        $\mathbf{P}(k;t)    =   [\mu(t)]^{-1}
                                                \big\{
                                                        \mathbf{I}-\mathbf{c}(k;t)[\mathbf{\hat{x}}_\mathcal{I}^{(t)}(k)]^H
                                                \big\}
                                                \mathbf{P}(k;t-1)$
                    \\
                    \IF{$t = N-1$}
                        \STATE $\mathbf{P}(k) = \mathbf{P}(k;N-1)$
                        \STATE $\mathbf{P}(k+1;0) = \mathbf{P}(k;N-1)$
                    \ENDIF
                    \\

                \ENDFOR
                \STATE
                    $\mathbf{w}(k+1)        =       \mathbf{P}(k)\mathbf{R}_{\mathcal{S}}(k)
                                                    \frac{
                                                            \displaystyle \mathbf{w}(k)
                                                        }
                                                        {
                                                            \displaystyle
                                                            \left\|\mathbf{w}(k)\right\|
                                                        }$
                \\

                \STATE $y_o(k) = \mathbf{w}^H(k) \mathbf{x}_\mathcal{S}(k)$
                \\

            \ENDFOR
        \end{scriptsize}
    \end{algorithmic}
\end{algorithm}

\section{Simulation Results and Discussions}
\label{Sec:Simulation} In this section, we provide numerical
examples to verify the validity of the proposed MIC-MPB scheme, and
compare the performance of it with that of the PAPC and the Maximin
beamformer. In the simulations, we assume the transmitted DPSK
signal is spreaded by a distinct $31$-chip Gold sequence ($N=31$)
and modulated onto carrier frequency of $1$ GHz for each user. The
data-symbol and spreading sequences are randomly generated for each
simulation trial at the rates of $100$ kbps and $3.1$ Mbps,
respectively. Since each signal path of the desired user is
processed separately by employing the two-dimensional RAKE receiver,
without loss of generality, we assume the desired user has one
propagation path in the first two subsections. In the last
subsection, we will discuss performance of the proposed beamformer
in a special case for RAKE processing, i.e., there are multipaths
with identical delay of the desired user.

\subsection{Ability of Suppressing
Structured Interference} \label{Sec:Simulation:Structured
Interference} Firstly, we study the ability of suppressing
structured interference of the beamformers. Three typical
scenarios--the received SOI with periodically repeated white noise,
tones, and MAIs in multipath channels are simulated with some
specially selected simulation parameters of the interferers. In all
the cases, we consider a uniform linear array (ULA) with eight
omnidirectional antennas ($L=8$) spaced half a wavelength apart. In
these simulations, we also assume that the SOI always arrives from
$0^{\circ}$ and the power of the interferers are always assumed to
be equal in each scenario.

\begin{figure}[!t]
\centering
\includegraphics[width=0.48\textwidth]{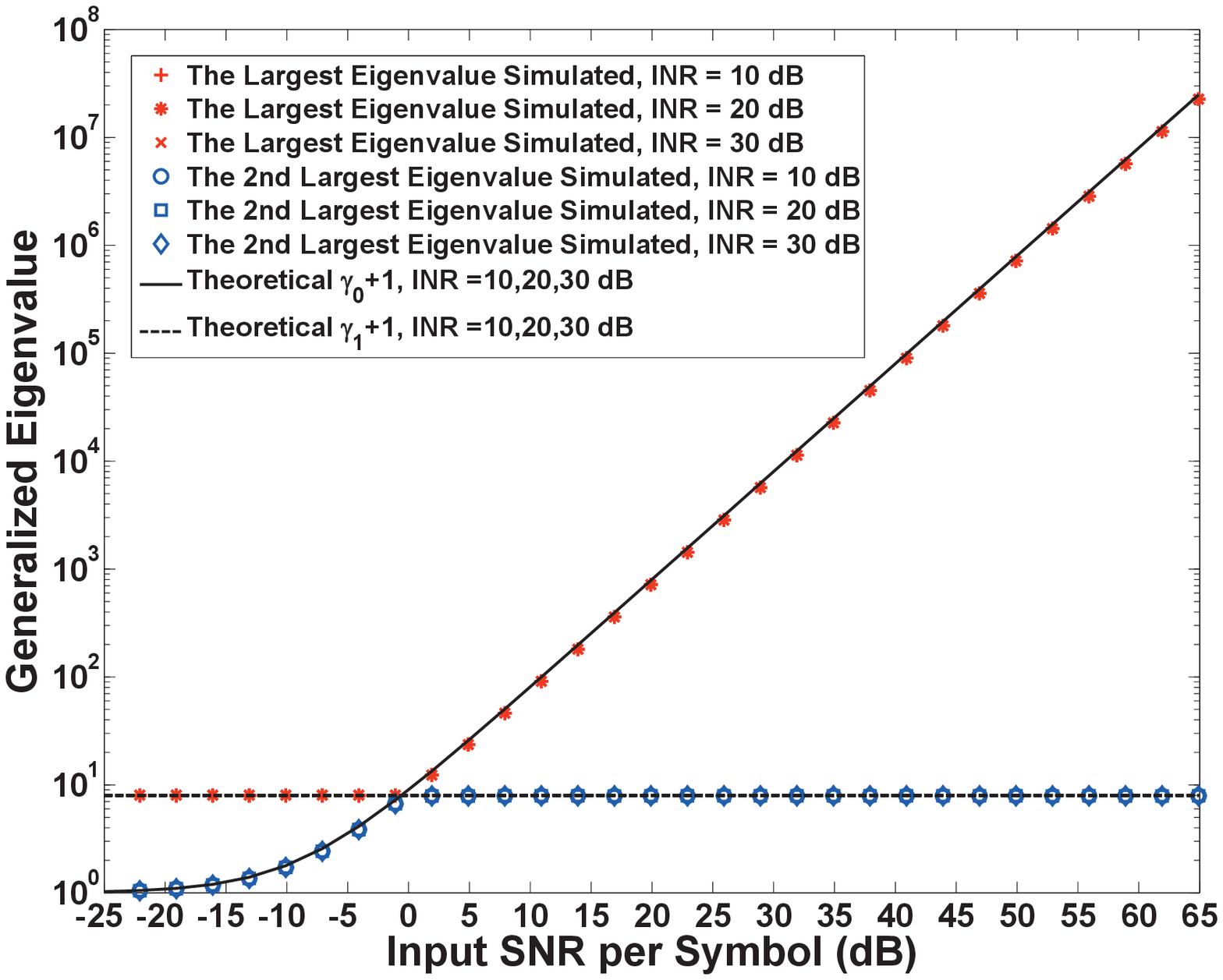}
\caption{The largest and 2nd largest generalized eigenvalues of the
MIC-MPB vs. $\textsf{SNR}$ in five tones case.}
\label{Fig:Simulation:MIC-Eigenvalue}
\end{figure}

Fig. \ref{Fig:Simulation:MIC-Eigenvalue} shows the largest and
second largest generalized eigenvalues of the matrix pair of the
MIC-MPB with five tones interferers. The tones are assumed to
impinge on the array from the directions $30^{\circ}$,
$-50^{\circ}$, $-20^{\circ}$, $19^{\circ}$, and $45^{\circ}$ with
frequency offsets $100$ kHz, $-300$ kHz, $0$, $400$ kHz, and $-100$
kHz, respectively, with respect to the carrier frequency of $1$ GHz
of the SOI. The simulated eigenvalues are obtained by computing the
matrix pair $\mathbf{R}_{\mathcal{S}}$ and
$\mathbf{R}_{\mathcal{I}}$ from generated received array signals
then using eigen-decomposition operation. In order to avoid finite
sample effects, $1$ million data symbols ($K=10^6$) are used to
estimate the covariance matrix pair. Theoretical $\gamma_{0}+1$ is
computed by (\ref{Equ:MPBeamformer:Gamma0}) and $\gamma_{1}+1$ by
using eigen-decomposition of the matrix pair
$(\mathbf{Q}_\mathcal{S},\mathbf{Q}_\mathcal{I})$. From this figure,
we can observe that when $\textsf{SNR}\leq -0.6$ dB,
$\gamma_{0}+1<\gamma_{1}+1$ and the largest eigenvalue of the matrix
pair equals $\gamma_{1}+1$; when $\textsf{SNR}> -0.6$ dB,
$\gamma_{0}+1$ linearly increases while $\gamma_{1}+1$ remains a
constant, the largest eigenvalue then switches to $\gamma_{0}+1$.
Therefore, the threshold of the MIC-MPB can be considered as $-0.6$
dB. Since $\gamma_{1}+1$ of the beamformer remains the same when the
power of the interferers or the interference-to-noise ratio (INR)
increases, the threshold of the MIC-MPB is small and bounded in this
scenario.

\begin{figure}[!t]
\centering
\includegraphics[width=0.48\textwidth]{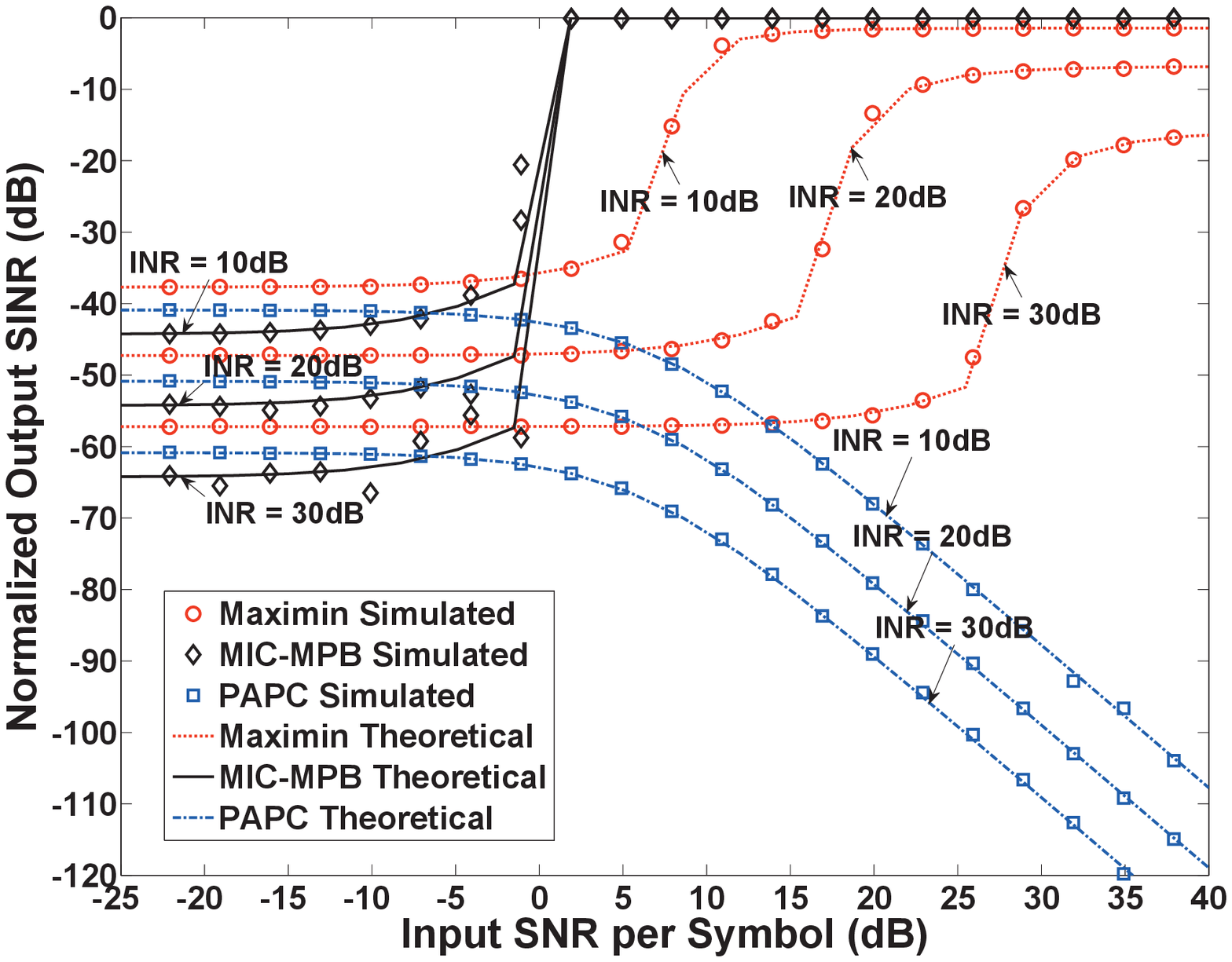}
\caption{Normalized output SINRs corresponding to the MIC-MPB,
Maximin, and PAPC vs. $\textsf{SNR}$ in two periodically repeated
white noise case.} \label{Fig:Simulation:SINRPN_vs_SNR}
\end{figure}

\begin{figure}[!t]
\centering
\includegraphics[width=0.48\textwidth]{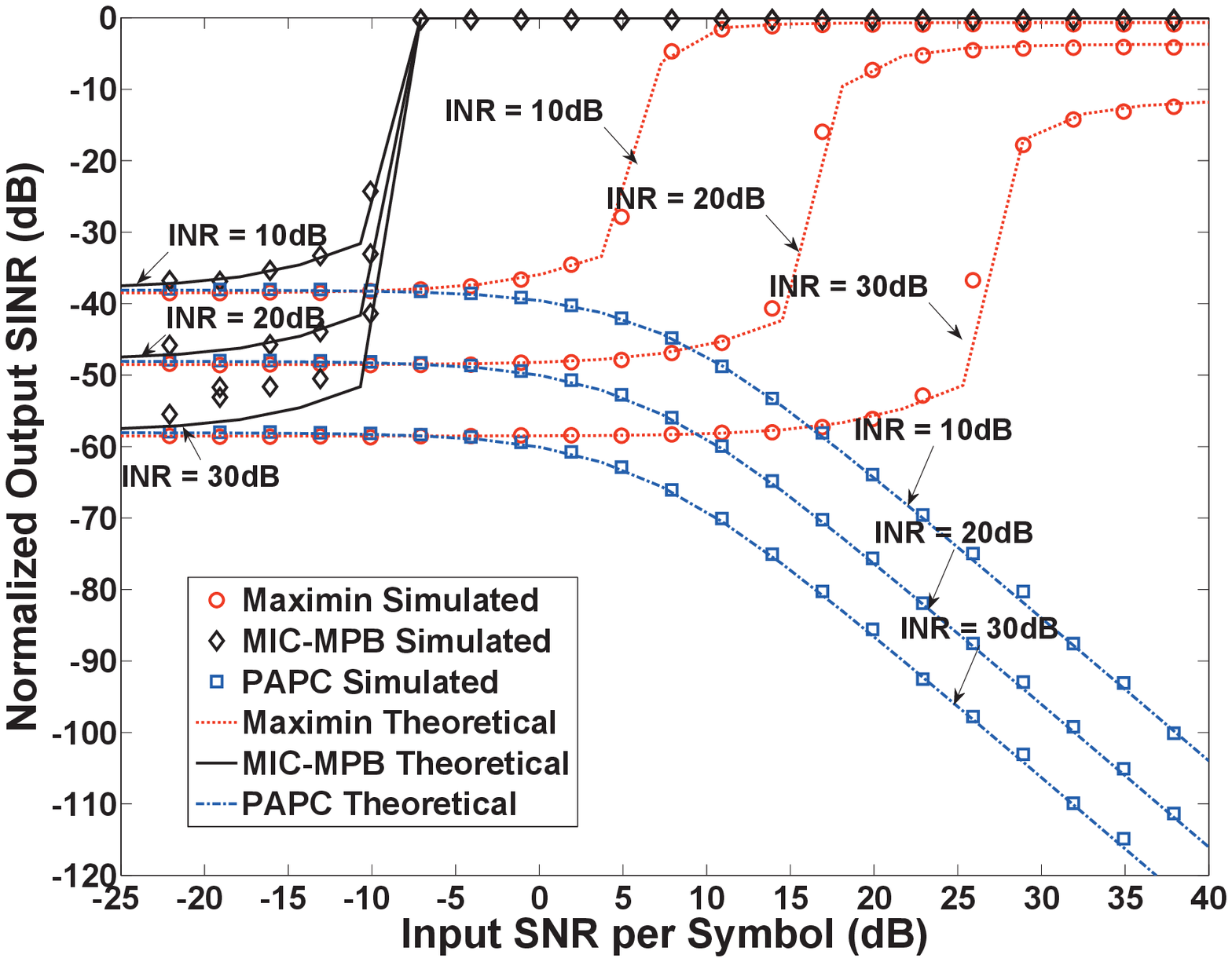}
\caption{Normalized output SINRs corresponding to the MIC-MPB,
Maximin, and PAPC vs. $\textsf{SNR}$ in three-ray multipath MAI
case.} \label{Fig:Simulation:SINRTM_vs_SNR}
\end{figure}

\begin{figure}[!t]
\centering
\includegraphics[width=0.48\textwidth]{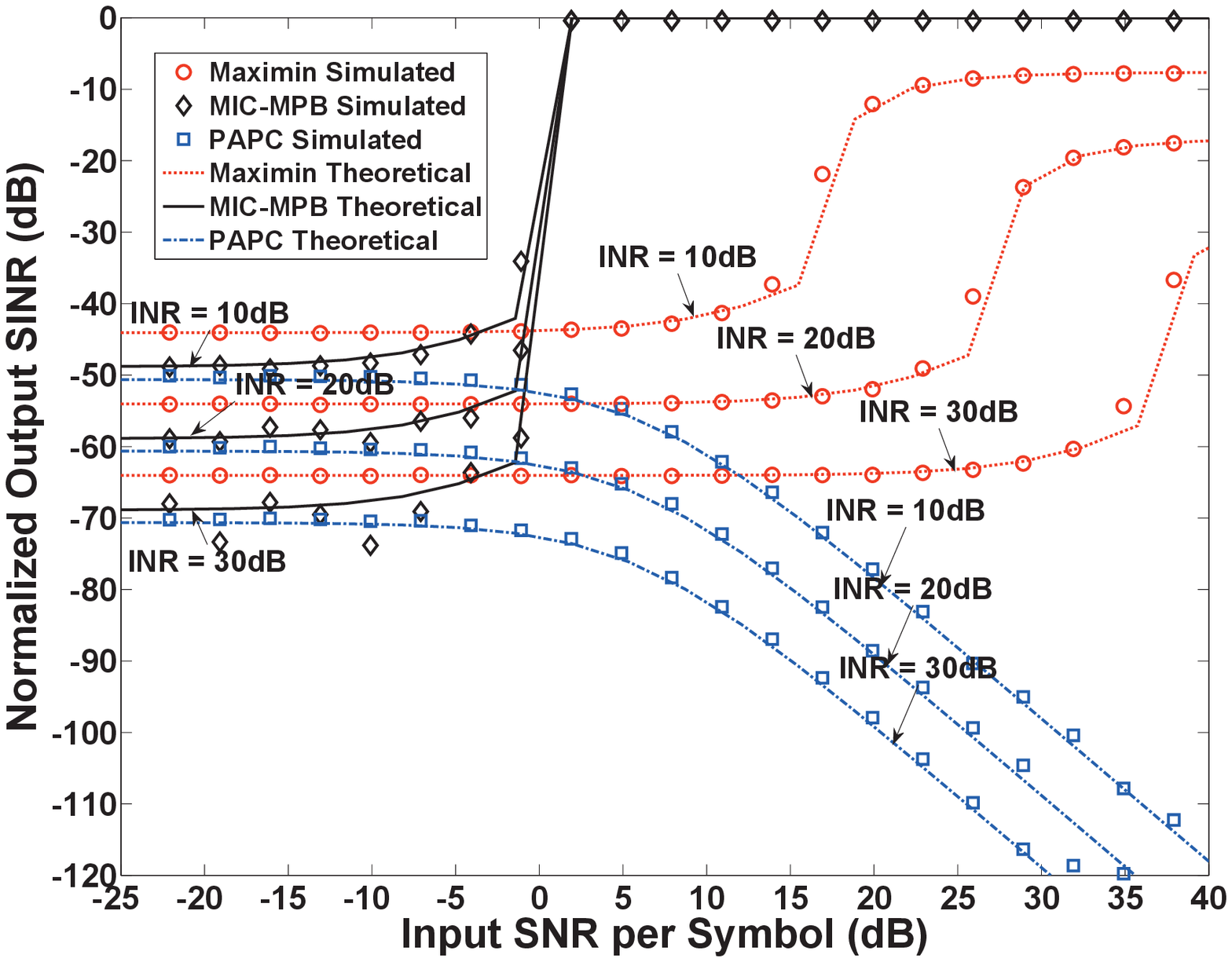}
\caption{Normalized output SINRs corresponding to the MIC-MPB,
Maximin, and PAPC vs. $\textsf{SNR}$ in five tones case.}
\label{Fig:Simulation:SINRTT_vs_SNR}
\end{figure}

Fig. \ref{Fig:Simulation:SINRPN_vs_SNR}--Fig.
\ref{Fig:Simulation:SINRTT_vs_SNR} show the normalized output SINRs
corresponding to the MIC-MPB, Maximin, and PAPC scheme versus input $\textsf{SNR}$ in the three scenarios. The normalized
output SINR is defined as the output SINR of the MPB
normalized by the optimum SINR with no interference, given by
\begin{align}
        \label{Equ:Simulation:G_init}
        \textsf{G} \triangleq \frac{\textsf{SINR}_{\mathrm{o}}}{\textsf{SINR}_{\mathrm{opt}}},
    \end{align}
where
\begin{align}
        \textsf{SINR}_{\mathrm{o}} & \triangleq  \frac{E\big\{|y_S(k)|^2\big\}}{E\big\{|y_I(k)|^2\big\}+E\big\{|y_N(k)|^2\big\}},    \nonumber\\
        \textsf{SINR}_{\mathrm{opt}} &=\frac{P_0}{\sigma^2}\cdot\|\mathbf{a}_0\|^2\cdot\|\mathbf{c}_0\|^2=L\textsf{SNR}.   \nonumber
    \end{align}

The simulated normalized output SINRs are obtained by using the
above equations with simulated received signals, and the theoretical
values are computed by an approximated piecewise function
$\textsf{G}(\textsf{SNR})$ described in \cite{Jianshu,Jianshu2}. In Fig.
\ref{Fig:Simulation:SINRPN_vs_SNR}, two periodically repeated white
noise arrive at $30^\circ$ and $-40^\circ$, respectively. The
periods of the interferers are both equal to the duration of a CDMA
symbol $T_{s}$. In Fig. \ref{Fig:Simulation:SINRTM_vs_SNR}, there is
one incident MAI signal with three-ray multipath delays of $3$
chips, $5$ chips, and $4$ chips from directions $30^\circ$,
$-20^\circ$, and $-50^\circ$, respectively. The simulation
parameters in Fig. \ref{Fig:Simulation:SINRTT_vs_SNR} are the same
as those in Fig. \ref{Fig:Simulation:MIC-Eigenvalue}. Some points
need to be noted that these simulation parameters are specially
designed in order to give prominence to the threshold effects the MPBs, because the threshold of the Maximin or PAPC is very
small (far more less than $\textsf{SNR}$) and the beamformers can be
well-behaved in most cases. Since $\textsf{G}$ reflects the limiting
performance of a beamformer, $K=10^{6}$ symbols are simulated for
each $\textsf{SNR}$ under given INRs in every experiment to
eliminate finite sample effects. However, deviation in simulated
values still can be seen in the figures when INR $=30$ dB and
$\textsf{SNR}$ are below the thresholds of the proposed MIC-MPB
scheme. This phenomenon can be explained by
(\ref{Equ:MPBeamformer:wopt}), i.e., when $\textsf{SNR}$ is below
the threshold, the steering vectors of the interferers will dominate
and the beamformer can be considered as an Miminum Power
Distortionless Response (MPDR) beamformer, which will receive the
interferers. Since larger INR means more interference power
contained in $\mathbf{R}_\mathcal{I}$, more data samples are
required for ``satisfactory'' performance
\cite{Trees2002,Chang1992}. But for the Maximin or PAPC beamformer,
things are totally different. This is because both schemes employ
one dimensional interference subspace $\mathcal{I}$, which make
independent interferers correlated after projection operation. As a
result, the steering vector of the interferers contained in
$\mathbf{R}_\mathcal{I}$ is a compound vector, which is different
from $\mathbf{a}_{\epsilon_1}$. Therefore, they can be considered as
MVDR beamformers when $\textsf{SNR}$ are below the thresholds, and
far more less samples are needed to maintain stable system
performance.

\begin{table}[!tb]
\renewcommand{\arraystretch}{1.3}
\caption{Input SNR Thresholds of the beamformers in two periodically
repeated white noise case} \label{Tab:Simulation:PNThreshold}
\begin{center}
\begin{tabular}{c|c|c|c}
\hline \hline  Matrix Pair & \multicolumn{3}{c}{Input SNR Thresholds $\textsf{SNR}_{\textsf{T}0}$ (dB)}\\
\cline{2-4} Beamformers  & INR $=10$ dB & INR $=20$ dB & INR $=30$ dB\\
\hline
MIC-MPB & $-0.93$ & $-0.85$ & $-0.84$ \\
\hline
Maximin & $7.7$ & $17.5$ & $27.5$\\
\hline
PAPC & $\infty$ & $\infty$ & $\infty$\\
\hline
\end{tabular}
\end{center}
\end{table}

\begin{table}[!tb]
\renewcommand{\arraystretch}{1.3}
\caption{Input SNR Thresholds of the beamformers in three-ray
multipath MAI case} \label{Tab:Simulation:TMThreshold}
\begin{center}
\begin{tabular}{c|c|c|c}
\hline \hline  Matrix Pair & \multicolumn{3}{c}{Input SNR Thresholds $\textsf{SNR}_{\textsf{T}0}$ (dB)}\\
\cline{2-4} Beamformers  & INR $=10$ dB & INR $=20$ dB & INR $=30$ dB\\
\hline
MIC-MPB & $-9.4$ & $-9.3$ & $-9.3$ \\
\hline
Maximin & $6.2$ & $15.8$ & $25.8$\\
\hline
PAPC & $\infty$ & $\infty$ & $\infty$\\
\hline
\end{tabular}
\end{center}
\end{table}
\begin{table}[!tb]
\renewcommand{\arraystretch}{1.3}
\caption{Input SNR Thresholds of the beamformers in five tones case}
\label{Tab:Simulation:TThreshold}
\begin{center}
\begin{tabular}{c|c|c|c}
\hline \hline  Matrix Pair & \multicolumn{3}{c}{Input SNR Thresholds $\textsf{SNR}_{\textsf{T}0}$ (dB)}\\
\cline{2-4} Beamformers  & INR $=10$ dB & INR $=20$ dB & INR $=30$ dB\\
\hline
MIC-MPB & $-0.64$ & $-0.56$ & $-0.55$ \\
\hline
Maximin & $16.4$ & $26.4$ & $36.4$\\
\hline
PAPC & $\infty$ & $\infty$ & $\infty$\\
\hline
\end{tabular}
\end{center}
\end{table}

\begin{figure}[!t]
\centering
\includegraphics[width=0.48\textwidth]{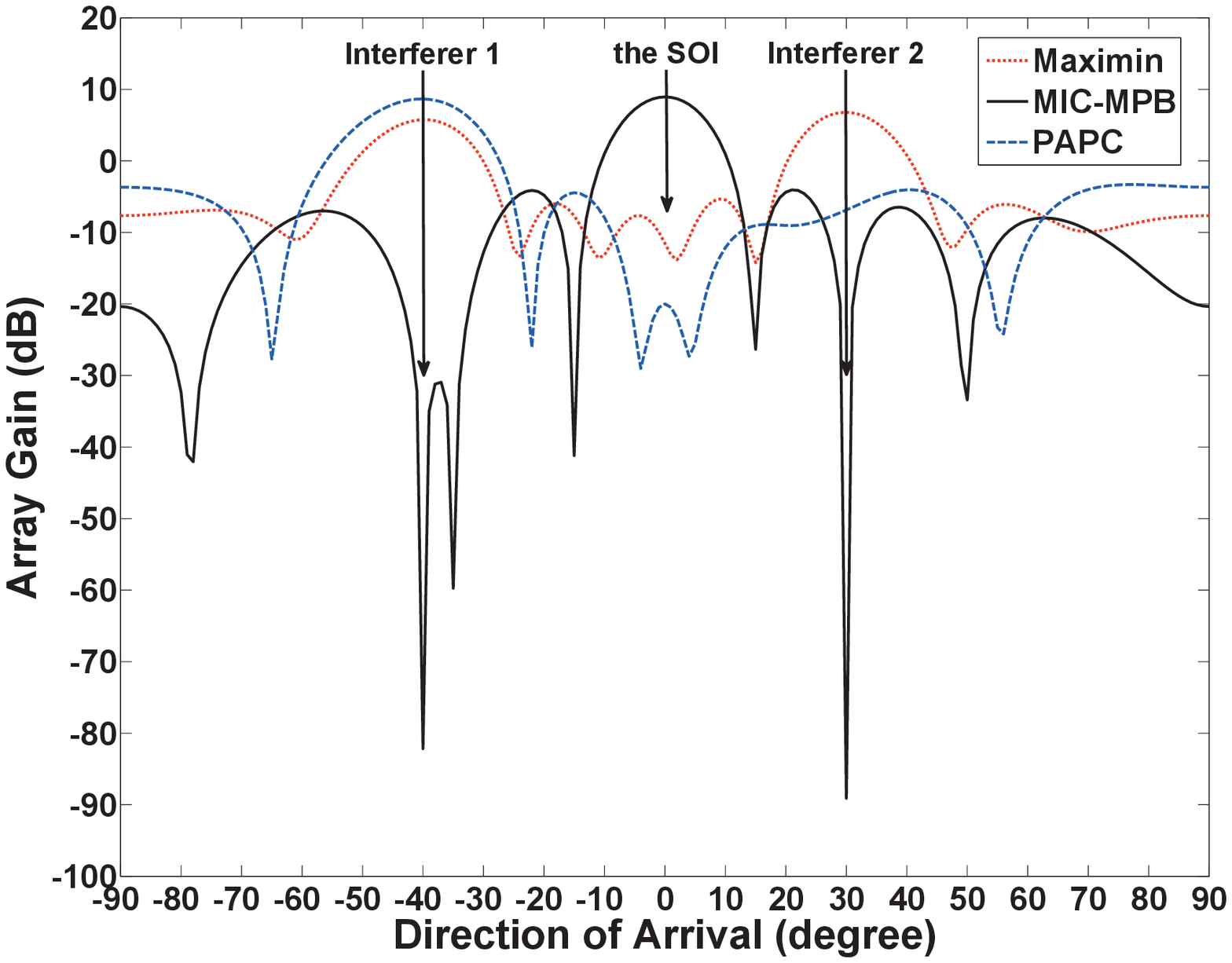}
\caption{The array patterns corresponding to the MIC-MPB, Maximin
and PAPC with $\textsf{SNR} = 40.9$ dB and INR $=30$ dB in two
periodically repeated white noise case.}
\label{Fig:Simulation:AP_Comp2}
\end{figure}

From the figures, we can find that the proposed MIC-MPB scheme can
achieve the optimum SINR regardless of the received power of
interference in the three scenarios when
$\textsf{SNR}>\textsf{SNR}_{\textsf{T}0}$, which means the
structured interference have been totally filtered under this
condition. But for the Maximin beamformer, more input signal power
is needed for it to reach the upper plateau when the power of the
interferers or INRs increase. Meanwhile, its limiting performance
decreases when INR grows. This is because the Maximin beamformer
cannot perfectly eliminate the interferers in these scenarios, which
can be verified by Fig. \ref{Fig:Simulation:AP_Comp2}, the Maximin
beamformer does not form deep nulls in the direction of the
interferers. For the PAPC beamformer, we can find that it completely
fails in the scenarios. Furthermore, its normalized output SINR
decreases to zero in the order of $\mathcal{O}(\textsf{SNR}^{-2})$
when $\textsf{SNR}$ goes to infinity.

Table \ref{Tab:Simulation:PNThreshold}--Table
\ref{Tab:Simulation:TThreshold} give the input SNR thresholds of the
beamformers in the three scenarios. From
(\ref{Equ:MPBeamformer:Gamma0}), the input SNR thresholds can be
determined as the following equation
\begin{equation}
        \label{Equ:Simulation:threshold}
        \textsf{SNR}_{\textsf{T}0}   =  \frac{N}{L}\cdot \frac{\gamma_{1}}{N-\beta(1+\gamma_{1})}.
\end{equation}
The values of the thresholds given in the tables are in accord with
what are shown in the corresponding figures in the same scenarios.
The thresholds of the proposed MIC-MPB scheme are far more less than
those of the Maximin or PAPC scheme, and remain constants when INRs
increase. The thresholds of the Maximin beamformer increase the same
amount accordingly when INRs increase $10$ dB. The thresholds of the
PAPC beamformer also show its failure because the values are always
infinity in the three scenarios.

\begin{figure}[!t]
\centering
\includegraphics[width=0.48\textwidth]{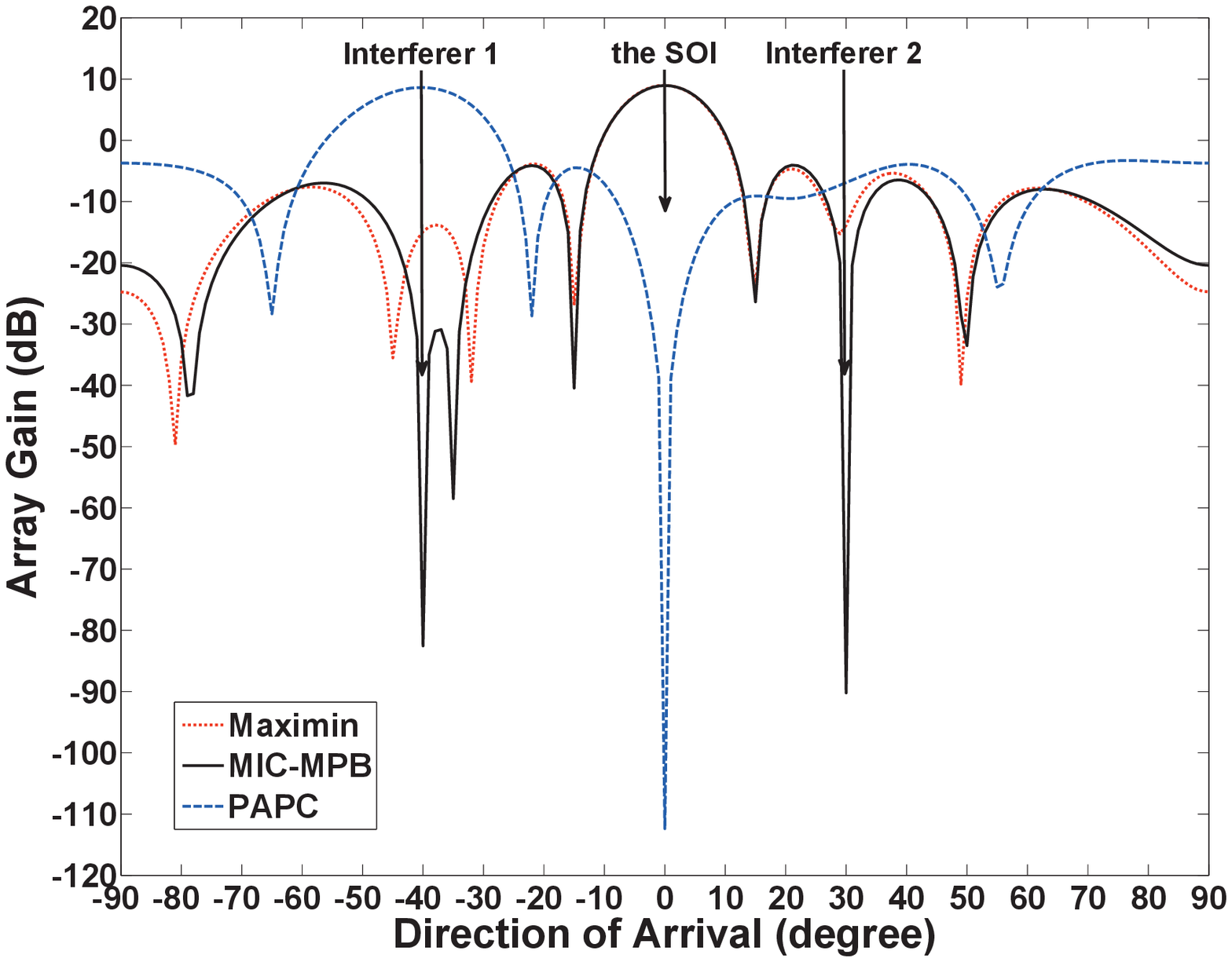}
\caption{The array patterns corresponding to the MIC-MPB, Maximin
and PAPC with $\textsf{SNR} = 10.9$ dB and INR $=30$ dB in two
periodically repeated white noise case.}
\label{Fig:Simulation:AP_Comp1}
\end{figure}

Fig. \ref{Fig:Simulation:AP_Comp1} and Fig.
\ref{Fig:Simulation:AP_Comp2} demonstrate the array patterns of the
MIC-MPB, Maximin, and PAPC beamformer in the two periodically
repeated white noise case. In Fig. \ref{Fig:Simulation:AP_Comp1},
the proposed MIC-MPB scheme can correctly receive the SOI and null
the interferes, but the Maximin or PAPC beamformer receives the
interferers and forms a side-lobe in the direction of the SOI. The
figure indicates that the MIC-MPB works at the operating area while
both the Maximin and PAPC beamformer work at the failure area for
$\textsf{SNR} = 10.9$ dB and INR $=30$ dB (c.f. Fig.
\ref{Fig:Simulation:SINRPN_vs_SNR} and Table
\ref{Tab:Simulation:PNThreshold}). In Fig.
\ref{Fig:Simulation:AP_Comp2}, the received signal power is very
large and $\textsf{SNR} = 40.9$ dB is much larger than
$\textsf{SNR}_{\textsf{T}0}$ of the MIC-MPB and Maximin algorithm,
so both algorithms can work properly. However, the Maximin
beamformer just form a side-lobe or a shallow notch in the direction
of the interferers. For the PAPC beamformer, a very deep null are
placed in the direction of the SOI for $\beta\neq 0$ and
$\mathbf{R}_{\mathcal{I}}$ contains part of the SOI, which can
partly explain why $\textsf{G}$ decreases when $\textsf{SNR}$ increases shown
in the above figures.

\begin{figure}[!t]
\centering
\includegraphics[width=0.48\textwidth]{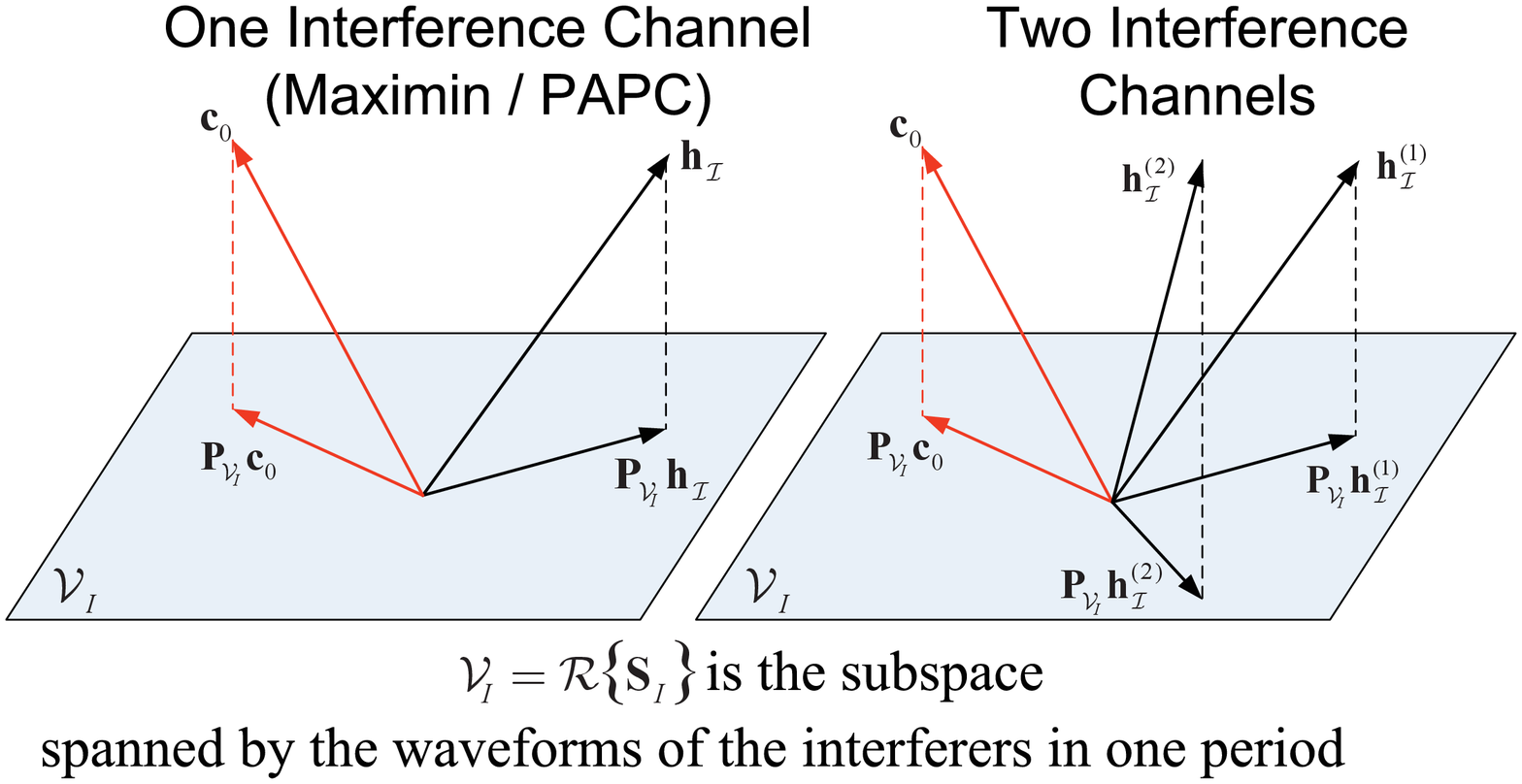}
\caption{Geometrical interpretation of different thresholds of the
MIC-MPB, Maximin and PAPC beamformer.}
\label{Fig:Simulation:illuChannels}
\end{figure}

Geometrical interpretation of different thresholds of the MIC-MPB,
Maximin and PAPC beamformer in the scenarios can be illustrated by
Fig. \ref{Fig:Simulation:illuChannels}. For the beamformer with one
interference channel or one dimensional interference subspace
$\mathcal{I}$, the condition
(\ref{Equ:MPBeamformer:SpaceCondition2b}) which make $\gamma_1$
bounded is equivalent to the condition that requires the projected
vectors of $\mathbf{H}_{\mathcal{I}}$ and $\mathbf{h}_{\mathcal{S}}$
onto $\mathcal{V}_{I}$ must be in one line (c.f. Fig.
\ref{Fig:Simulation:illuChannels}). But this condition can hardly be
satisfied for uncertainty of the characteristics of the interferers.
For the proposed beamforming scheme with multiple interference
channels, since there are multiple base vectors in the interference
channel, the condition can be easily satisfied.

\subsection{Performance of convergence rate with finite samples}
\label{Sec:Simulation:FiniteSamples}

\begin{figure}[!t]
\centering
\includegraphics[width=0.48\textwidth]{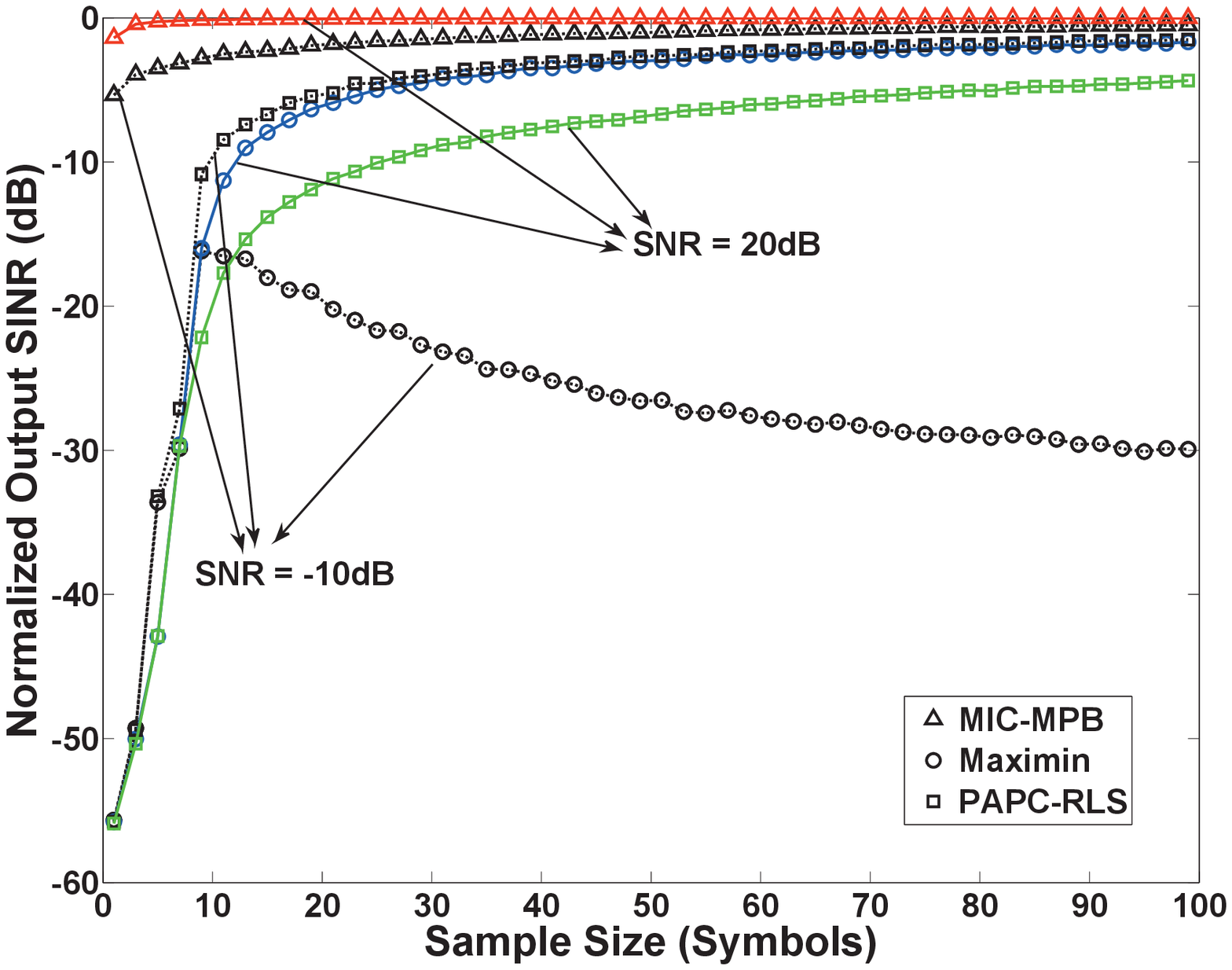}
\caption{Normalized average output SINR corresponding to the
MIC-MPB, Maximin, and PAPC-RLS vs. sample size under various
$\textsf{SNR}$ in 1000 trials.} \label{Fig:Simulation:SINRout_vs_K}
\end{figure}

In this subsection, we compare the performance of convergence rate
of the MPBs with finite samples. In the simulations, we
assume the receiver has an array of ten elements ($L=10$) with half
wavelength spacing, and receives a single path SOI from $20^\circ$.
There are seven MAIs, with INR of $40$ dB and DOAs of $35^\circ$,
$-35^\circ$, $-45^\circ$, $0^\circ$, $-50^\circ$, $-60^\circ$ and
$45^\circ$, respectively. Moreover, a broadband BPSK jamming also
arrives from $60^\circ$ with INR of $40$ dB. These parameters have
been verified not to cause obvious threshold effects of the Maximin
and PAPC beamformer. Since there are two different
approaches-stochastic gradient method \cite{choi2002nab} and
recursive least squares (RLS) method \cite{Yang2006fbba} for PAPC
beamformer to search the optimal weight vector in the literature, we
name the algorithms as PAPC-SG and PAPC-RLS respectively for
notational convenience. Fig. \ref{Fig:Simulation:SINRout_vs_K} shows
the normalized output SINRs, defined as the ratio of output SINRs to
the optimum value $\textsf{SINR}_{\mathrm{opt}}$ under given $\textsf{SNR}$,
which are calculated by averaging over $1000$ independent trials. We
observe that the proposed MIC-MPB scheme converges to the optimum
performance within a few symbols, and is independent of the desired
signal strength. In contrast, the PAPC-RLS and Maximin schemes
require much more symbols and the performance of PAPC-RLS degrades
when the input $\textsf{SNR}$ increases. These
results confirm the performance improvement of the MIC-MPB scheme,
which extracts more effective samples per data symbol and eliminates
the desired component in interference subspace $\mathcal{I}$.


\begin{figure}[!t]
\centering
\includegraphics[width=0.48\textwidth]{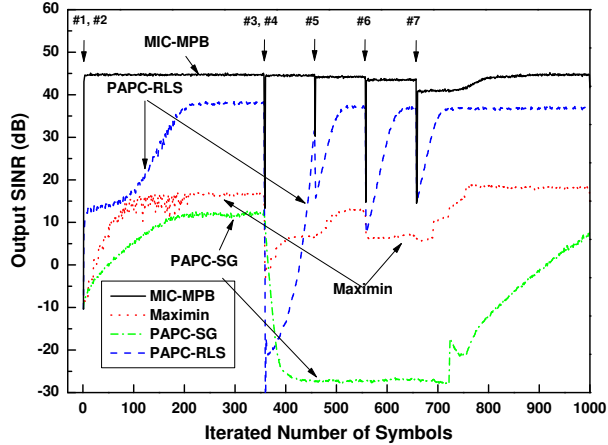}
\caption{Output SINR vs. time of the four beamformers in dynamic
multiple access channel in 1000 trials.}
\label{Fig:Simulation:Dynamic_Performance}
\end{figure}

We also simulate the performance of different adaptive algorithms
for dynamic multiple access channels. In this simulation, the input $\textsf{SNR}$ is fixed to $20$ dB. Fig.
\ref{Fig:Simulation:Dynamic_Performance} compares the tracking
ability of the PAPC-SG, PAPC-RLS, Maximin algorithm and MIC-MPB
algorithm presented in section \ref{Sec:MIC-MPB:RecursiveAlgorithm}.
The DOAs of the seven MAIs are identical to the previous simulation.
The first two MAIs are $8$ dB stronger than the power of the SOI and
the others are $40$ dB stronger. The time they enter the channel are
marked in the figure. The results demonstrate that the proposed
recursive algorithm can null the new interferers within a few
symbols, much faster than the other three algorithms.

%

\subsection{Performance when there are multipaths with identical delays}
\label{Sec:Simulation:angular} 
In practice, the scatterers local to the mobile will cause an
angular spread of about $3^\circ$ at a distance of $1$
km\cite{LAL1997}, and the relative delays between the multipaths are
generally small. Thus, the assumption that the relative delays are
greater than one chip may not hold. In this subsection, we will show
that the proposed beamformer still work well under such condition.

Assume there are $D_{i}$ paths for the $i$th user. We first define a
set $\mathcal{U}_i \triangleq
\{1,2,\ldots,D_i\}=\bigcup_{s=1}^{S_i}\mathcal{U}_{i,s}$, so that
the subset $\mathcal{U}_{i,s}$ satisfies
\begin{enumerate}
\item $\forall s \neq s',\,\mathcal{U}_{i,s} \cap \mathcal{U}_{i,s'} =
\varnothing$;
\item $\forall j,j' \in \mathcal{U}_{i,s},\,n_{ij} = n_{ij'} =
n_{is}$.
\end{enumerate}
where $n_{ij}$, $n_{ij'}$, and $n_{is}$ all denote the equivalent
propagation delays of certain paths. Thus, $\mathcal{U}_{i,s}$
contains all the $i$th user's path indices of the same delay. As a
result, we can rewrite (\ref{Equ:Signal_model:received_discrete}) as
\begin{align}
\label{Equ:Simulation:signalmodel_IDdelay} \mathbf{x}(n) &=
\sum_{i=0}^{M-1} \sum_{s=1}^{S_i} \sum_{k=-\infty}^{+\infty} b_i(k)
c_i(n-n_{is}-kN) \mathbf{\tilde{a}}(\theta_{is})+ \sum_{q=1}^{Q} z_q(n) \mathbf{a}(\theta_{q}) + \mathbf{v}(n)
\end{align}
where $\mathbf{\tilde{a}}(\theta_{is}) \triangleq \sum_{j \in
\mathcal{U}_{i,s}} \sqrt{P_{ij}}\mathbf{a}(\theta_{ij})$ is the
compound steering vector. For the desired user ($i=0$) and $\forall
j \in \mathcal{U}_{0,s}$, the matrices $\mathbf{R}_{\mathcal{S}}$
and $\mathbf{R}_{\mathcal{I}}$ will only depend on $s$, so we denote
them as $\mathbf{R}_{\mathcal{S},s}$ and
$\mathbf{R}_{\mathcal{I},s}$ respectively. The $s$th beamformer is
then
\begin{equation}
\label{Equ:Simulation:w_opts_IDdelay} \mathbf{w}_{\mathrm{opt},s} =
\mu \mathbf{R}_{\mathcal{I},s}^{-1}\mathbf{\tilde{a}}(\theta_{0s}) =
\mu \sum_{j \in \mathcal{U}_{0,s}} \sqrt{P_{0j}}
\mathbf{R}_{\mathcal{I},s}^{-1} \mathbf{a}(\theta_{0j}),
\end{equation}
which means that the $s$th beamformer will cancel all other signals
except the ones having the delay of $n_{0s}$. Moreover, multiple
beams will be formed to collect and combine the multipath components
from different directions. Therefore, the algorithm is still
applicable in such situation, and the only variation is that just
$S_0$ beamformers are required.

\begin{figure}[!t]
\centering
\includegraphics[width=0.48\textwidth]{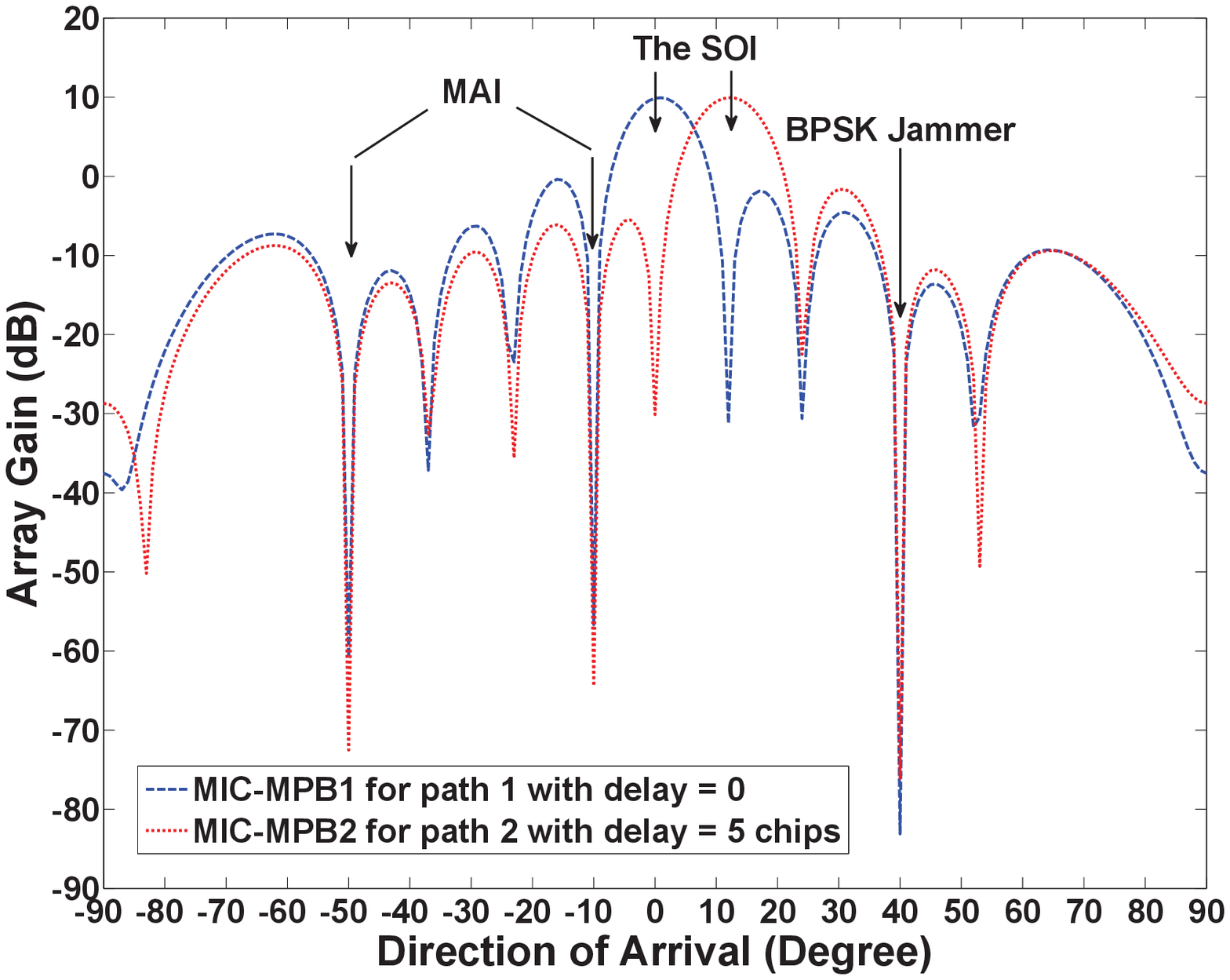}
\caption{Array patterns of MIC-MPB for paths
with different delays.}
\label{Fig:Simulation:Angular_Spread2}
\end{figure}
\begin{figure}[!t]
\centering
\includegraphics[width=0.48\textwidth]{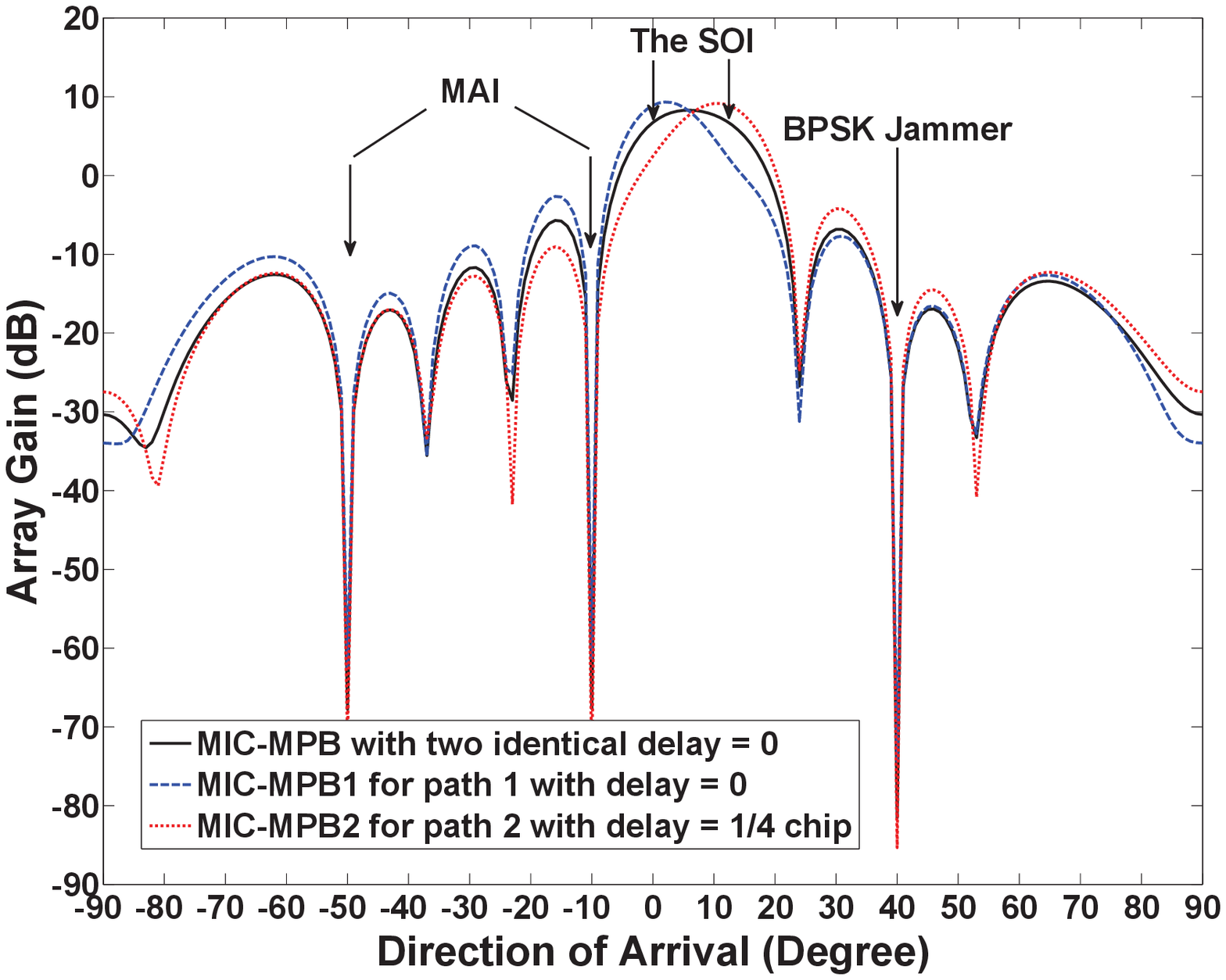}
\caption{Array patterns of MIC-MPB for paths
with different and identical delays.}
\label{Fig:Simulation:Angular_Spread1}
\end{figure}

Fig. \ref{Fig:Simulation:Angular_Spread2} and Fig.
\ref{Fig:Simulation:Angular_Spread1} show the simulated array
patterns when the delays are different (dash line and dot lines),
and the array pattern when the delays are identical (solid line). In
the simulation, array elements $L = 10$ with half wavelength spacing
are considered. Two users ($M=2$) communicates with the receiver.
The first user is the desired one and the second user acts as an
MAI. There is a BPSK jammer from $40^\circ$ and INR $=40$ dB. The
bandwidth of the broadband jammer is $1/T_{c}$. Each user has two
paths with equal power. The DOAs of the two desired paths are
$0^\circ$ and $12^\circ$. The paths of the second user arrive from
$-10^\circ$ and $-50^\circ$, and are $20$ dB stronger than each path
of the desired user. The input $\textsf{SNR}$ for
each desired path is $15$ dB.
In the former situation, the
proposed MIC-MPB scheme forms two different beams to collect the two
paths respectively, and each beamformer will suppress the other path
besides the MAIs and the jammer. If the two desired paths have the
identical delay, then one uniform beam will be formed to receive
them, only nulling the MAIs and the jammer. Fig. \ref{Fig:Simulation:Angular_Spread1} also shows when delays are not discriminable within one-chip period, two different beams will still be formed, but the two desired path are both collected by each beam. This implies that the
proposed approach is robust to angular spread, where the delay spread is small.

\section{Conclusion}
In this paper, we presented the principles for designing the
projection space which are closely correlated with the ability of
suppressing structured interference and system finite sample
performance. According to the principles, we proposed an MIC-MPB
scheme for CDMA systems which can be efficiently implemented by FFT.
We also derived an adaptive algorithm for the beamformer.
Computation and simulation results show that the proposed beamformer
has a small and bounded SNR threshold, and can achieve the optimum
SINR regardless of the received power of interference in the
scenarios with structured interference. Furthermore, the various
simulation results illustrate that the proposed MIC-MPB scheme has
better finite sample performance, faster convergence rate and more
superior tracking capability in the dynamical environment than the
existing MPBs.


\ifCLASSOPTIONcaptionsoff
  \newpage
\fi

\end{document}